\documentclass[12pt]{article}
\pdfoutput=1
\usepackage{amssymb,amsmath}
\usepackage{mathtools}
\usepackage{graphicx}
\usepackage{epsfig}
\usepackage{epstopdf}
\usepackage{setspace}
\usepackage{latexsym}
\usepackage{tikz}
\usepackage{psfrag}
\usepackage{booktabs}
\usepackage{bbm}
\usepackage[toc]{appendix}
\usepackage{color}
\usepackage{datetime}
\usepackage[
      colorlinks=true,
      linkcolor=darkblue,  
      urlcolor=blue,    
      filecolor=blue,     
      citecolor=red,
      linktocpage=true,
      pdfstartview=FitV,
      bookmarksopen=true
      ]{hyperref}

\usepackage{tikz}
\usetikzlibrary{arrows,shapes}
\usetikzlibrary{trees}
\usetikzlibrary{patterns}
\usetikzlibrary{matrix,arrows} 				% For commutative diagram
\usetikzlibrary{positioning}				% For "above of=" commands
\usetikzlibrary{calc,through}				% For coordinates
\usetikzlibrary{decorations.pathreplacing}  % For curly braces
\usepackage{pgffor}							% For repeating patterns

\usetikzlibrary{decorations.pathmorphing}	% For Feynman Diagrams
\usetikzlibrary{decorations.markings}
\tikzset{
    vector/.style={decorate, decoration={snake}, draw},
	provector/.style={decorate, decoration={snake,amplitude=2.5pt}, draw},
	antivector/.style={decorate, decoration={snake,amplitude=-2.5pt}, draw},
        smallvector/.style={decorate, decoration={snake,amplitude=1.5pt,post length=0.5mm}, draw},
    fermion/.style={draw=black, postaction={decorate},
        decoration={markings,mark=at position .55 with {\arrow[draw=black]{>}}}},
    fermionbar/.style={draw=black, postaction={decorate},
        decoration={markings,mark=at position .55 with {\arrow[draw=black]{<}}}},
    fermionnoarrow/.style={draw=black},
    gluon/.style={decorate, draw=black,
        decoration={coil,amplitude=4pt, segment length=5pt}},
    scalar/.style={dashed,draw=black, postaction={decorate},
        decoration={markings,mark=at position .55 with {\arrow[draw=black]{>}}}},
    scalarbar/.style={dashed,draw=black, postaction={decorate},
        decoration={markings,mark=at position .55 with {\arrow[draw=black]{<}}}},
    scalarnoarrow/.style={dashed,draw=black},
    electron/.style={draw=black, postaction={decorate},
        decoration={markings,mark=at position .55 with {\arrow[draw=black]{>}}}},
    bigvector/.style={decorate, decoration={snake,amplitude=4pt}, draw},
    arrow/.style={draw=black, postaction={decorate},
        decoration={markings,mark=at position 1 with {\arrow[draw=black]{>}}}},
}

\tikzstyle{block} = [draw, rectangle, 
    minimum height=3em, minimum width=6earticlem]

\definecolor{darkblue}{rgb}{0.2, 0, 0.8}

\numberwithin{equation}{section}

\usepackage{cite,dsfont}
\usepackage{amsfonts}
\usepackage{amssymb}
\usepackage{amsmath}
\usepackage{graphicx}
\usepackage{slashed}
\usepackage{setspace}

\usepackage{color}

\newcommand{\reef}[1]{(\ref{#1})}

\newcommand{\be}{\begin{equation}}
\newcommand{\ee}{\end{equation}}

\def\be{\begin{equation}}
\def\ee{\end{equation}}
\def\bea{\begin{eqnarray}}
\def\eea{\end{eqnarray}}
\def\ba{\begin{array}}
\def\ea{\end{array}}
\def\bd{\begin{displaymath}}
\def\ed{\end{displaymath}}

\def\>{\rangle}                     % right angle
\def\<{\langle}                     % left angle
\def\Dsl{D \hskip-.6em \raise1pt\hbox{$ / $ } }
\def\to{\rightarrow}

\newcommand{\eps}{\epsilon}
\newcommand{\lra}{\leftrightarrow}

\newcommand{\KLT}[1]{\stackrel{\mathclap{\mbox{\tiny{#1}}}}{\otimes}}

\begin{document}  
%%%%%%%%%%%%%%%%%%%%%%%%%%%%%%%%%%%%%%%

%%%%%%% title page %%%%%%%%%

\begin{titlepage}

 \begin{flushright}
{\tt LCTP-21-34} \\
\end{flushright}

\vspace*{2.3cm}

\begin{center}
  \setstretch{2.0}
{\LARGE \bf On Extended Supersymmetry of 4d Galileons and 3-Brane Effective Actions} \\
  \setstretch{1.0}
\vspace*{1.2cm}

\hspace{-0.14in}{\bf Henriette Elvang and Matthew Dominique Mitchell}
\medskip

Leinweber Center for Theoretical Physics,\\ 
Randall Laboratory of Physics, Department of Physics,\\
University of Michigan, Ann Arbor, MI 48109, USA

\bigskip
{\small elvang@umich.edu, mattdmit@umich.edu}  \\
\end{center}

\vspace*{0.1cm}

\begin{abstract}  

We use on-shell amplitude methods to systematically analyze the possibility of extended supersymmetry for 4d Galileon models, expanding on previous $\mathcal{N}=1$ results. 
Assuming spins $\le 1$, we prove that there exists no $\mathcal{N}=4$ supersymmetric extension of any 4d  Galileons. Thus the Galileons cannot be part of the effective action of a single flat maximally supersymmetric D3-brane, and that explains why such terms do not appear in the $\alpha'$-expansion of the abelian open superstring amplitude. 
For $\mathcal{N}=2$ Galileons,  we show that 
the complex scalar $Z = \phi + i \chi$ of the  vector supermultiplet cannot have $\phi$ and $\chi$ both enjoy enhanced shift symmetry; instead, $\chi$ can at best be an $R$-axion with constant shift symmetry. Using the soft bootstrap, we demonstrate that the quartic DBI-Galileon is incompatible with $\mathcal{N}=2$ supersymmetry. A similar analysis  performed at 7-point shows that a 2-parameter family of $\mathcal{N}=2$ supersymmetric quintic Galileons coupled with DBI passes the soft bootstrap. Finally, we show how supersymmetric coup\-lings between Galileons and gravitons arise in generalizations of our constructions, and we conclude with a discussion of      Galileons and DBI-Galileons in the context of UV-completability vs.~the Swampland.
\end{abstract}

\end{titlepage}

%%%%%%%%%%%%%%%%%%%%%%%%%%%%%%%%%%%%%
\setcounter{tocdepth}{3}
%the line above sets the depth of the table of contents. {2} means it will display section and subsections only.
{\small
\setlength\parskip{-0.5mm}
\tableofcontents
}

\newpage

%%%%%%%%%%%%%%%%%%%%%%%%%%%%%%
\section{Introduction}
\label{sec:Introduction}
%%%%%%%%%%%%%%%%%%%%%%%%%%%%%%

A flat brane in Minkowski space spontaneously breaks part of the spacetime symmetry: each broken translational symmetry \cite{Low:2001bw} gives rise to a massless Goldstone boson $\phi$ in the low-energy effective action on the brane and the broken rotation/boost symmetries induce an enhanced shift symmetry  \cite{deRham:2010eu,Goon:2010xh}
\be \label{shiftsym}
   \phi \to \phi + c + c_\mu x^\mu +\text{field-dependent terms}\,,
\ee
for constants $c$ and $c_\mu$.
The leading-order low-energy effective theory of these  Goldstone bosons is the Dirac-Born-Infeld model (DBI), here written in static gauge for a single scalar
\be
  \label{DBI}
  \mathcal{L}_\text{DBI} = 
  - \Lambda^4 
  \bigg( 
  \sqrt{-\det\big(\eta_{\mu\nu}
  +
  \tfrac{1}{\Lambda^4}
  \partial_\mu \phi\partial_\nu \phi\big)} -1
  \bigg)\,,
\ee
with $\Lambda^4$ the brane tension. 
It indeed has the shift symmetry \reef{shiftsym}. 

It is well-known that DBI has an $\mathcal{N}=4$ supersymmetric extension which describes the leading-order dynamics of a single D3-brane in 10d Minkowski space. The six massless real scalars of the $\mathcal{N}=4$ vector multiplet are the Goldstone modes of the 6 broken transverse translational symmetries, the four fermions are Akulov-Volkov (A-V) Goldstinos of the partially-broken 10d supersymmetry, and the vector particle is a Born-Infeld photon. 
$\mathcal{N}=4$ supersymmetric DBI has several nice properties, including the scalar shift symmetry \reef{shiftsym}, fermionic shift symmetry of the A-V Goldstinos, global $SO(6) \sim SU(4)$ symmetry, and electromagnetic duality.

Now consider higher-derivative corrections to DBI. In a general bottom-up Effective Field Theory (EFT)  approach, the admissible higher-derivative operators are the local scalar operators compatible with the enhanced shift symmetry \reef{shiftsym}. For a single real scalar in 4d, the {\em cubic, quartic, and quintic Galileon} interaction terms 
\bea
   \label{gal3action}
   \mathcal{L}_{\text{Gal}_3}
   &=&  
  \frac{g_3}{2}\, (\partial \phi)^2 \Box \phi
  \,, \\[1mm]
 \label{gal4action}
   \mathcal{L}_{\text{Gal}_4}
   &=&  
  \frac{g_4}{3}\, (\partial \phi)^2
  \big(
     (\Box \phi)^2 
     - (\partial\partial\phi)^2
  \big)\,, \\[1mm]
   \label{gal5action}
   \mathcal{L}_{\text{Gal}_5}
   &=&\frac{g_5}{3}\, 
   (\partial \phi)^2
  \big(
     (\Box \phi)^3 
     - 2(\partial\partial\phi)^3
     + 3(\Box \phi)(\partial\partial\phi)^2
  \big)
  \,
\eea
are the lowest-order operators beyond DBI that realize these shift symmetries. When combined with a canonical kinetic term, the equation of motion of the Galileons  \reef{gal3action}-\reef{gal5action} are second order despite their higher-derivative origin in the Lagrangian and they are free of Ostrogradski ghosts \cite{Nicolis:2008in}. For these reasons, Galileons are considered attractive, and they arise in various contexts: the DGP model \cite{Dvali:2000hr}, modified gravity \cite{Nicolis:2008in,deRham:2010ik,deRham:2010kj}, and --- as noted above --- as candidates for higher-derivative terms in low-energy brane effective actions \cite{deRham:2010eu}.

The cubic interaction is equivalent to a linear combination of the quartic and quintic Galileons via a field redefinition, so we do not consider it further and instead we focus our attention on the quartic and quintic Galileons. Further, while from a pure low-energy point of view, Galileons can be considered in their own right, it is only in the presence of DBI that (at least the quartic) Galileon can be a UV-completable EFT \cite{Adams:2006sv,Arkani-Hamed:2020blm}. Of course including DBI is also natural from the perspective of the effective 3-brane action. The combination of DBI and Galileons is called {\em DBI-Galileons}.

The context of brane effective actions and the $\mathcal{N}=4$ supersymmetry of DBI are primary motivations for examining if Galileons too are compatible with supersymmetry. Earlier studies have largely focused on $\mathcal{N}=1$ supersymmetry. 
The authors of 
\cite{Koehn:2013hk} examined examples of 4d $\mathcal{N}=1$ cubic Galileons, but they did not find any that were ghost-free. 
In \cite{Farakos:2013zya}, a 4d $\mathcal{N}=1$ supersymmetric quartic Galileon model was constructed using superfields. 
The first systematic analysis of $\mathcal{N}=1$ supersymmetry for the 4d quartic and quintic Galileons was performed in  \cite{Elvang:2017mdq,Elvang:2018dco} using on-shell amplitudes methods. Evidence was found in favor of $\mathcal{N}=1$ supersymmetrization of  both the quartic and quintic Galileons; the key results of  \cite{Elvang:2017mdq,Elvang:2018dco} are reviewed in Section \ref{s:review}. Recently, the superalgebras underlying supersymmetric  EFTs with (enhanced) shift symmetries, such as the Galileons, were studied in \cite{Roest:2019dxy}.

{\em The main goal of this paper is to perform a systematic analysis of extended supersymmetry of 4d Galileon models. Specifically we examine compatibility of the 4d quartic and quintic Galileons with $\mathcal{N}=4$ and $\mathcal{N}=2$ supersymmetry with and without the presence of DBI.}

In string theory, higher-derivative corrections to DBI can be calculated from the $\alpha'$-expansion of the abelian open string tree-level scattering  amplitudes \cite{Tseytlin:1987ww,Andreev:1988cb,Tseytlin:1999dj}. For the bosonic string, the leading corrections  at 4-point enter at  6-derivative order, i.e.~the order of the quartic Galileon.  This can be seen in the 4-vector amplitudes \cite{Tseytlin:1987ww,Andreev:1988cb}, but also directly in the 4-scalar amplitudes.\footnote{We thank Stephan Stieberger for communications about this point.} For the superstring, corrections are postponed until 8-derivative order \cite{Tseytlin:1999dj}; this means that the superstring does {\em not} produce the quartic Galileon in the $\alpha'$-expansion of the D3-brane effective action. 
We show in this paper that the reason simply is that the 4d quartic Galileon is not compatible with $\mathcal{N}=4$ supersymmetry.

In our analysis of extended supersymmetry, we make the following assumptions:
\begin{itemize}
\item The models are local, unitary, and they have 4d Poincar\'e symmetry. 
\item All spins are $\le 1$ and there are no 3-point interactions. (In Section \ref{s:SG} we relax these assumptions.)
\item The on-shell particle spectrum is organized into massless supermultiplets:
\begin{itemize}
\item  $\mathcal{N}=1$ SUSY: chiral supermultiplet,
\item  $\mathcal{N}=2$ SUSY: vector-multiplet,\footnote{One could also consider the $\mathcal{N}=2$ hypermultiplet, but we do not pursue that in this paper.} and
\item  $\mathcal{N}=4$ SUSY: the CTP self-conjugate  vector supermultiplet. 
\end{itemize}
\item There is at least one real Galileon $\phi$ with shift symmetry \reef{shiftsym} in the spectrum and it has to sit in one of the complex scalars $Z$ in the supermultiplet. Writing $Z = \phi + i \chi$, there are three options to consider for $\chi$:
\begin{itemize} 
\item[(a)] $\chi$ is also a Galileon that realizes the enhanced shift symmetry \reef{shiftsym}, or
\item[(b)] 
 $\chi$ is an $R$-axion with only constant shift symmetry, or
\item[(c)]  $\chi$ has no shift symmetry.\label{casec}
\end{itemize}
\item The spin-1/2 fermions in the multiplet are Goldstino-like in the sense that they have a shift symmetry $\lambda \to \lambda + \xi$, where $\xi$ is a constant Grassmann-valued spinor. 
\end{itemize}

Part of our analysis actually does not require the assumption that $\chi$ or the fermions  have any  shift symmetry. In other cases, option (b) and the shift symmetry of the fermions emerges from the imposed of supersymmetry and assumption that $\phi$ is a Galileon.

To be considered a supersymmetrization of the Galileon, the tree amplitudes with only real Galileons as the external states should be exactly equal to those of the real scalar Galileon models. This requirement turns out not to constrain the constructions;  models that pass the other constraints, automatically have this property.

Our analysis takes advantage of on-shell amplitudes methods to examine the space of possible 4d Galileon models with extended  supersymmetry. 
The 4- or 5-point on-shell matrix elements of the quartic and quintic models, respectively, must be polynomial in the 4d spinor helicity spinor-brackets; they cannot have poles because that would require them to factorize into 3-point amplitudes of which we assume there are none. This makes it simple to systematically characterize all possible 4- or 5-point matrix elements. Next, these are subjected to the supersymmetric Ward identities of $\mathcal{N}$-fold SUSY as well as constraints of (enhanced) soft theorems, i.e.~Adler zeroes, associated with the (enhanced) shift symmetries. For any models that pass at 4- and 5-point, we use the soft bootstrap   \cite{Cheung:2014dqa,Kampf:2013vha,Cheung:2015cba,Cheung:2015ota,Cheung:2016drk}\cite{Elvang:2018dco}, reviewed in Appendix \ref{app:softboot}, to test consistency. This allows us to very easily rule out models --- or find evidence (though not proof) of their existence. 

In Section \ref{s:review} we review the shift symmetries and associated soft behavior of scalars and fermions, We also provide an overview of the results of \cite{Elvang:2017mdq,Elvang:2018dco} for $\mathcal{N}=1$ supersymmetry of the quartic and quintic Lagrangian. We present a new compact formula for the complex scalar amplitude $A_5(\bar{Z} Z \bar{Z} Z \bar{Z})$ of the 3-parameter family of $\mathcal{N}=1$ compatible quintic Galileon model. 

Below we summarize the results of our analysis. Assuming spins $\le 1$:
\begin{itemize}
\item {\bf $\mathcal{N}=4$ supersymmetry} \\[1mm]
In  Section \ref{s:noN4Gal}, we use locality and supersymmetric Ward identities  to prove that there exists no $\mathcal{N}=4$ supersymmetrizations of the quartic or quintic Galileons.

\item {\bf $\mathcal{N}=2$ supersymmetry}\\[1mm]
Section \ref{s:N2gal} is dedicated to $\mathcal{N}=2$ supersymmetry. We show 
\begin{itemize}
\item in Section \ref{s:N2SUSY4ptsigmaZ2} that there does not exist an $\mathcal{N}=2$ quartic Galileon in which both $\phi$ and $\chi$ are Galileons (option a); one of the real scalars can at best be an $R$-axion (option b). On its own, such an  $\mathcal{N}=2$ quartic Galileon-axion model cannot be constrained by the soft bootstrap. 
\item in Section \ref{s:N2Gal4sigma1} that the $\mathcal{N}=2$ quartic Galileon model with $\chi$ an $R$-axion or no shift symmetry at all (options b or c) cannot exist in the presence of supersymmetric DBI: it is excluded by the soft bootstrap assuming  a constant shift symmetry of the fermions. This rules out $\mathcal{N}=2$ for quartic DBI-Galileon. 
  \item in Section \ref{s:N2quinticsigmaZ1} that there is evidence for a 
   $\mathcal{N}=2$ quintic Galileon in which the Galileon $\phi$ is joined by an $R$-axion (option b).\footnote{Option (a) with two Galileons was ruled out for the $\mathcal{N}=1$ quintic Galileon \cite{Elvang:2017mdq,Elvang:2018dco}, as we review in Section \ref{s:N1Gal}.} It passes all soft bootstrap tests at 7-point when combined with $\mathcal{N}=2$ DBI. 
\end{itemize}
\end{itemize}
Note: we do not claim that any of the resulting candidates for supersymmetric Galileon models are ghost-free. The analysis is focused on the constraints of supersymmetry and soft behavior (i.e.~shift symmetry).

In the above analysis there were no 3-point interactions. If we allow them, we are forced to include particles with spin greater than 1, and in particular gravitons show up:
\begin{itemize}
\item {\bf Galileons and Gravity}\\[1mm]
Allowing the 4-point amplitudes to have poles, the $\mathcal{N}\ge 2$ supersymmetry Ward identities  predict a coupling between the Galileon vector superpartner and a graviton. We briefly examine a few such Galileon-supergravity scenarios with $\mathcal{N}=2$ and $\mathcal{N}=4$ supersymmetry in Section \ref{s:SG}.
In some of these models, the couplings between supergravity and Galileons discussed in Section \ref{s:SG}  involve massless higher-spin states and as such they are not expected to be UV-completable \cite{Hinterbichler:2017qcl}.
\item  {\bf Galileons and Swampland}\\[1mm]
 We review the positivity arguments \cite{Adams:2006sv,Arkani-Hamed:2020blm} that in the absence of DBI, the quartic Galileon is in the Swampland. We then consider the bounds on the quartic Galileon coupling from the S-matrix bootstrap \cite{Caron-Huot:2020cmc,Bellazzini:2020cot,Tolley:2020gtv,Guerrieri:2020bto} and discuss them in perspective of our supersymmetry results. 
\end{itemize}

We conclude in Section \ref{s:discussion} with a discussion of brane constructions of (DBI-)Galileons  as well as the absence of supersymmetric Galileons in  double-copy constructions. 

There are three appendices. Appendix \ref{app:softboot} contains a concise review of the soft bootstrap in the form used in this paper and offers some details of the soft bootstrap calculations from the main text.  Appendix \ref{app:N1quintic} outlines the derivation the 3-parameter $\mathcal{N}=1$ SUSY quantic Galileon. Appendix \ref{app:SUSYWI} derives an $\mathcal{N}=4$ SUSY Ward identity used in Section \ref{s:noN4Gal}.

%%%%%%%%%%%%%%%%%%%%%%%%%%%%%
\section{Review}
\label{s:review}
%%%%%%%%%%%%%%%%%%%%%%%%%%%%%

In this section we describe the connection between shift symmetries and vanishing soft theorems and we review the results of \cite{Elvang:2017mdq,Elvang:2018dco} for 4d $\mathcal{N}=1$  Galileons.

\subsection{Vanishing Soft Theorems}
In the soft limit of a single massless particle,  
the tree amplitude behaves as
\be  
   \label{softAn}
   A_n \to \mathcal{O} \left( \eps^\sigma \right)
   ~~~~\text{with}~~~~
   p_s^\mu \to \eps p_s^\mu 
   ~~~\text{as}~~~
   \eps \to 0 \,,
\ee
for some integer {\em soft weight} $\sigma$. 
The soft limit $p_s^\mu \to \eps p_s^\mu =  -\eps |s\> [s|$ is implemented on the spinor helicity variables as follows:
\be
  \label{softconv}
  \begin{split}
  |s\> &\to \eps |s\> ~~~\text{for}~~h_s \ge 0\,,\\
  |s] &\to \eps |s] ~~~\,\text{for}~~h_s < 0 \,,
  \end{split}
\ee
where $h_s$ is the helicity of the soft particle. 
This way of taking the soft limit ensures that trivial soft factors of $\eps$ from little group scaling are eliminated. 

In the absence of cubic interactions, we have:
\begin{equation}
\label{softness}
\text{
\begin{tabular}{lc}
Shift symmetry & Soft Weight $\sigma$\\
$\phi \to \phi + c + \ldots$ & 1 \\
$\phi \to \phi + v_\mu x^\mu + \ldots$ & 2 \\
$\phi \to \phi + s_{\mu\nu} x^\mu x^\nu + \ldots$ & 3 \\
$\psi \to \psi + \xi + \ldots$ & 1 \\
\end{tabular}}
\end{equation}
where $s_{\mu\nu}$ is traceless \cite{Hinterbichler:2015pqa} and $\xi$ is a constant Grassmann-valued spinor. The ``$+\ldots$" stand for potential field redefinitions. For more details of the relationship between shift symmetries and soft behavior, see for example \cite{Cheung:2016drk,Elvang:2018dco}.

In the on-shell approach to effective field theories, sometimes called the soft bootstrap, the desired shift symmetries of the model are imposed in the form of these soft theorems on the on-shell scattering amplitudes. In particular, {\em we take soft weight $\sigma=2$ for a real scalar to be the defining feature of a Galileon.}

%%%%%%%%%%%%%%%%
\subsection{4d Galileons and $\mathcal{N}=1$ SUSY.}
\label{s:N1Gal}
The matrix elements of the quartic and quintic Galileon interactions \reef{gal4action} and \reef{gal5action} are
\be
  \label{realGalAs}
  A_4(\phi\phi\phi\phi) = g_4\,  s t u\,,~~~~
  A_5(\phi\phi\phi\phi\phi) = g_5\, \big(\eps(1,2,3,4)\big)^2\,,
\ee
where the Levi-Civita contraction is 
\be
  \label{eps1234}
  \eps(1,2,3,4) 
  \equiv p_{1\mu} p_{2\nu} p_{3\rho} p_{4\lambda} \eps^{\mu\nu\rho\lambda}
  = \<12\>[23]\<34\>[41]
    -[12]\<23\>[34]\<41\>\,.
\ee
 The $n$-point amplitudes of the quartic Galileon have an extra-enhanced soft behavior with soft weight $\sigma=3$ \cite{Cheung:2014dqa} due to an additional shift symmetry quadratic in the position \cite{Hinterbichler:2015pqa}, as in the third line of \reef{softness}. 
When the Galileon is interacting with other particles, the special enchancement to $\sigma=3$ is lost and the soft weight of the Galileon is $\sigma=2$.

In a supersymmetrization, the real Galileon scalar field must be part of a complex scalar field 
\be
  \label{defZ}
  Z = \frac{1}{\sqrt{2}}(\phi + i\chi) \,, ~~~~
  \bar{Z} = \frac{1}{\sqrt{2}}(\phi - i\chi) \,. 
\ee
Assuming that $\phi$ is a Galileon, i.e.~$\sigma_\phi = 2$, the options for $\chi$ are:
\begin{enumerate}
\item[(a)] $\chi$ is also a Galileon with enhanced soft weight $\sigma_\chi =2$; or
\item[(b)]  $\chi$ is an $R$-axion with softness  $\sigma_\chi =1$. 
\end{enumerate}
Option (a) gives the complex scalar softness $\sigma_Z = \sigma_{\bar{Z}}=2$ while option (b) reduces it to $\sigma_Z = \sigma_{\bar{Z}}=1$.

Option (c) from page \pageref{casec}, where $\chi$ has no  shift symmetry would correspond to $\sigma_\chi \le 0$. 
Negative $\sigma_\chi$ requires cubic interactions which we are not considering here. Softness $\sigma_\chi =0$ is logically an option, but we find that $\sigma_\chi =1$ emerges from the $\sigma_\phi =2$ constraints at 4- and 5-point and we need only impose the softness of $\chi$ in a few cases when we use soft recursion to higher points. 

\subsubsection{$\mathbf{\mathcal{N}=1}$ Quartic Galileon} 
The most general 4-point complex scalar amplitude with two $Z$'s and two $\bar{Z}$'s at 6th order in momentum (which is what we need for the Galileon) takes the  form
\be
  \label{preA4ZZbZZb}
  A_4(Z \bar{Z} Z \bar{Z}) = c_1 s t u + c_2 t^3\,.
\ee
These are the only two independent Mandelstam polynomials of degree 3 compatible with Bose symmetry. 

The soft bootstrap, reviewed in Appendix \ref{app:softboot}, can be used to determine if the ansatz \reef{preA4ZZbZZb} is compatible with the soft behavior of options (a) and (b). As shown in Appendix \ref{app:n4nosusy}, the 6-point soft bootstrap gives 
\be
  \label{4ptsigmaZ2}
  \text{option (a): }\sigma_Z=\sigma_{\bar{Z}} =2 ~~\text{requires}~~ c_2 = 0]\,.
\ee
and 
\be
  \label{4ptsigmaZ1}
  \text{option (b): } \sigma_Z=\sigma_{\bar{Z}} =1 ~~\text{is compatible with}~~ c_1, c_2 \ne 0\,.
\ee

For both $c_1$ and $c_2$ non-zero, the 4-point amplitude \reef{preA4ZZbZZb} is compatible with $\mathcal{N}=1$ supersymmetry \cite{Elvang:2017mdq}. 
For the case $c_2 = 0$, the 4- and 6-point amplitudes (computed from soft recursion and SUSY Ward identities) were shown to \cite{Elvang:2017mdq} match those constructed directly from the superspace Lagrangian of the $\mathcal{N}=1$ quartic Galileon proposed in  \cite{Farakos:2013zya}. The case of $c_1 = 0$ is relevant for our discussion of $\mathcal{N}=2$ supersymmetry in Section \ref{s:N2gal}.

%%%%%%%

\subsubsection{$\mathbf{\mathcal{N}=1}$ Quintic Galileon} 
There are 10 independent local matrix element $A_5(Z \bar{Z} Z \bar{Z} Z)$ at order $O(p^8)$ compatible with Bose symmetry. Among these, there is one unique one compatible with $\sigma_Z=\sigma_{\bar{Z}} =2$ \cite{Elvang:2017mdq}, namely 
\be
\label{A5sigmaZ2}
A_5(Z \bar{Z} Z \bar{Z} Z) =  g_5' \,\big(\eps(1,2,3,4)\big)^2\,.
\ee
with $\eps(1,2,3,4)$ defined in \reef{eps1234}. 
However, this matrix element is {\em incompatible} with the 5-point $\mathcal{N}=1$ SUSY Ward identities \cite{Elvang:2017mdq}. This therefore rules out option (a) for a supersymmetric quintic Galileon.\footnote{This assumes absence of poles in the 5-point matrix elements; see the discussions in Sections \ref{s:SG}.}

It was found in \cite{Elvang:2017mdq,Elvang:2018dco}  that there is a 3-parameter family of local matrix elements  $A_5(\bar{Z}Z\bar{Z} Z \bar{Z})$ compatible with   $\mathcal{N}=1$ SUSY and the soft behavior of option (b), specifically
\be
  \sigma_\phi = 2,~~~\sigma_\chi = 1,~~~\sigma_\lambda = 1\,.
\ee
Appendix \ref{app:N1quintic} briefly reviews the 5-point analysis. Note that only  $\sigma_\phi=2$ is assumed, while $\sigma_\chi=1$ and $\sigma_\lambda=1$ emerges when $\mathcal{N}=1$ supersymmetry is imposed.

Let us now present a new simple form of the $\mathcal{N}=1$ 5-scalar  amplitude. It is
\bea
   \nonumber
  &&A_5^\text{$\mathcal{N}=1$ Gal$_5$}(\bar{Z}Z\bar{Z} Z \bar{Z})
  = q_1\,  s_{24}^2\Big[ s_{13}^2 +  s_{15}^2 +  s_{35}^2 - s_{24}^2\Big]
  \\
 \label{A5scalar3par}
  &&\hspace{1cm}+ q_2 \, s_{24} \bigg[ 
    - 2  \big( s_{13}^3 +  s_{15}^3 +  s_{35}^3\big)
    +    \sum_{i=1,3,5}\sum_{j=2,4} 
    \Big(2 s_{ij}^3+3s_{24} s_{ij}^2 \Big) \bigg]
   \\
   \nonumber
   &&\hspace{1cm}+q_3 \bigg[ 
   \frac{1}{2} \eps(1,2,3,4)
   \Big(
      (s_{12} - s_{14}) (s_{13} - s_{15})
      + \text{cyclic(1,\,3,\,5)}
   \Big)
   -3\eps(1,2,3,4)^2
   \bigg]\,,
\eea
where ``cyclic(1,\,3,\,5)" stands for adding the two terms with labels 1, 3, 5 cyclically permuted. 
The three parameters $q_1$, $q_2$, and $q_3$ are 
free. Projecting out the real scalars via \reef{defZ} gives
\be
  A_5^\text{$\mathcal{N}=1$ Gal$_5$}(\phi\phi\phi\phi\phi)= \frac{1}{\sqrt{2}}
  \big(
    2 q_1 + 12 q_2 - 15 q_3
  \big)\,
  \eps(1,2,3,4)^2\,,
\ee
which is indeed the real Galileon 5-point amplitude \reef{realGalAs}. Moreover, we find 
$A_5(\phi\phi\phi\phi\chi) = 0$, which ensures that the higher-point all-$\phi$ tree amplitudes obtained from soft recursion are those of the quintic Galileon because there are no possible $\chi$-exchanges. Finally, we find that $A_5(\chi\chi\chi\chi\chi) = 0$ which means that the $R$-axion has no quintic self-interactions.\footnote{As a curiosity, let us note that $A_5(\phi\phi\chi\chi\chi)$  depends only on $q_3$; in particular, this amplitude vanishes when $q_3=0$. The two remaining 5-scalar amplitudes 
$A_5(\phi\chi\chi\chi\chi)$ and $A_5(\phi\phi\phi\chi\chi)$ are generally non-zero and depends on all three parameters $q_1$, $q_2$, and $q_3$.} 

Let us now turn to the soft bootstrap. Going from two $\mathcal{N}=1$ 5-point Galileons to 8-point, there are only two constructible amplitudes: the  all-$\phi$ amplitude and  $A_8(\lambda\bar{\lambda}\phi\phi\phi\phi\phi\phi)$. Both pass the soft bootstrap. In combination with  leading-order $\mathcal{N}=1$ super-DBI, many amplitudes can be tested at 7-point via the soft bootstrap; see Appendix \ref{app:n5}. We have performed the soft bootstrap on all such  constructible DBI-Galileon 7-point amplitudes and found no restrictions on the 3 parameters of the $\mathcal{N}=1$ Galileon amplitude \reef{A5scalar3par}.\footnote{This contrasts the claim in \cite{Elvang:2017mdq,Elvang:2018dco} that the soft bootstrap reduces the 3-parameter family to a single unique solution. Upon re-examination, the previous work contained a parametrization error.}

\subsubsection{Summary}
To briefly summarize, we have the following candidates for 4d  $\mathcal{N}=1$ supersymmetric Galileon models \cite{Elvang:2017mdq,Elvang:2018dco}:
\begin{itemize}
\item A quartic Galileon with $\sigma_Z=2$ (two Galileon scalars), 
\item A 2-parameter family of quartic Galileons with $\sigma_Z=1$ (Galileon + $R$-axion), 
\item A 3-parameter family of quintic Galileons with $\sigma_Z=1$ (Galileon + $R$-axion).
\end{itemize}
We are now ready to examine the prospects if any of these models are compatible with extended supersymmetry.

%%%%%%%%%%%%%%%%%%%%%%%%%%%%%
\section{$\mathcal{N}=4$ Supersymmetry}
\label{s:noN4Gal}
%%%%%%%%%%%%%%%%%%%%%%%%%%%%%

In this section, we prove that the 4d quartic and quintic Galileons are not compatible with $\mathcal{N}=4$ supersymmetry. We assume that all 4- and 5-point matrix elements do not have any poles. 

%%%%%%%%%%%%%%%%%%%%%%%%%%%%%
\subsection{No Quartic $\mathcal{N}=4$ Galileon}
%%%%%%%%%%%%%%%%%%%%%%%%%%%%%

The massless $\mathcal{N}=4$ supermultiplet consists of a vector, four sets of fermions, and six real scalars that can be grouped into three complex scalars. We label the complex scalars $Z^{AB} = - Z^{BA}$ by the $SU(4)$ $R$-indices $A,B=1,2,3,4$ of the global $R$-symmetry of the $\mathcal{N}=4$ superalgebra, however we do not assume that the model realizes $SU(4)$. Suppose, without loss of generality, that the Galilon $\phi$ is the real part of the complex scalar $Z=Z^{12}$ and its conjugate  $\bar{Z}=Z^{34}$. The  $\mathcal{N}=4$ SUSY Ward identities then relate the 4-photon amplitude to the 4-scalar amplitude as
\be
  \label{vec2gal}
  A_4(\gamma^+ \gamma^- \gamma^+ \gamma^-) = \frac{\<24\>^2}{\<13\>^2} A_4(Z \bar{Z} Z  \bar{Z}).
\ee
Thus a $\mathcal{N}=4$ supersymmetrization of the quartic Galileon must contain the non-vanishing 4-vector amplitude \reef{vec2gal}.

The $\mathcal{N}=4$ vector supermultiplet is CTP self-conjugate, so the positive and negative helicity photon states are related by supersymmetry. In particular, the following $\mathcal{N}=4$ SUSY Ward identity must hold
\be
  \label{N4SUSYvec4pt}
  A_4(\gamma^- \gamma^- \gamma^+ \gamma^+) = 
  \frac{\<12\>^4}{\<24\>^4}
  A_4(\gamma^+ \gamma^- \gamma^+ \gamma^-)  \,.
\ee 
This Ward identity holds for the vector amplitudes of $\mathcal{N}=4$ supersymmetric DBI for which 
\be
 A_4(\gamma^- \gamma^- \gamma^+ \gamma^+) = \frac{1}{\Lambda^4}\<12\>^2[34]^2 \,
 ~~~\text{and}~~~
 A_4(\gamma^+ \gamma^- \gamma^+ \gamma^-) = \frac{1}{\Lambda^4}\<24\>^2[13]^2 \,.
\ee 
This is easily seen from
\be
   \label{BImani}
   \frac{\<12\>^4}{\<24\>^4}
   \frac{A_4(\gamma^+ \gamma^- \gamma^+ \gamma^-)}{A_4(\gamma^- \gamma^- \gamma^+ \gamma^+)} 
   = \frac{\<12\>^4}{\<24\>^4} \frac{\<24\>^2[13]^2}{\<12\>^2[34]^2} = 1 \,,
\ee
using 4-point momentum conservation $\<12\>[13] = - \<24\>[34]$. 

For the quartic Galileon, the matrix elements have mass-dimension 6 and --- by locality, little group scaling, and Bose symmetry --- that leaves only one option each for the local matrix elements with four external vectors, namely 
\be
 \label{4photon}
 A_4(\gamma^- \gamma^- \gamma^+ \gamma^+) = \frac{1}{\Lambda^6}\<12\>^2[34]^2 s \,
 ~~~\text{and}~~~
 A_4(\gamma^+ \gamma^- \gamma^+ \gamma^-) = \frac{1}{\Lambda^6}\<24\>^2[13]^2t \,.
\ee 
Now the same manipulations as in \reef{BImani} give  
\be
   \label{Gal4mani}
   \frac{\<12\>^4}{\<24\>^4}
   \frac{A_4(\gamma^+ \gamma^- \gamma^+ \gamma^-)}{A_4(\gamma^- \gamma^- \gamma^+ \gamma^+)} 
   = \frac{t}{s} \,.
\ee
Since $s \ne t$ for generic momenta, the amplitudes \reef{4photon} do not satisfy the $\mathcal{N}=4$ SUSY Ward identity \reef{N4SUSYvec4pt}. Hence, {\em the quartic Galileon is not compatible with $\mathcal{N}=4$ global supersymmetry.} Note that no assumptions were made about softness of any of the particles.

%%%%%%%%%%%%%%%%%%%%%%%%%%%%%
\subsection{No Quintic $\mathcal{N}=4$ Galileon}
\label{s:N4gGal5}
%%%%%%%%%%%%%%%%%%%%%%%%%%%%%

Suppose the Galileon is the real part of the complex scalar $Z = Z^{12}$ of an $\mathcal{N}=4$ supermultiplet. The conjugate scalar is then $\bar{Z} = Z^{34}$. Since the $\mathcal{N}=4$ supermultiplet is self-conjugate, supersymmetry relates $Z$ and $\bar{Z}$ to each other. In particular, we show in Appendix \ref{app:SUSYWI} that the following Ward identity has to hold in a model with $\mathcal{N}=4$ supersymmetry:
\be
  \label{5ZN4}
  \big\< \bar{Z} Z \bar{Z} Z  Z \big\>
  =
  -\frac{s_{13}}{s_{24}} \,
  \big\< Z \bar{Z} Z \bar{Z} Z \big\>\,.
\ee
The amplitude on the LHS is simply the RHS amplitude with momentum relabelings $(2 \lra 3, 1 \lra 4)$. Now plug the  3-parameter $\mathcal{N}=1$ solution \reef{A5scalar3par} into \reef{5ZN4}: one finds that the only solution is $q_1 = q_2 = q_3 = 0$. {\em Hence the quintic Galileon is incompatible with supersymmetry.}\footnote{One could more generally test the 10-parameter family of matrix elements that do not satisfy any softness constraint. In that case, the identity \reef{5ZN4} gives a unique solution, but it is incompatible with  $\sigma_Z=1$.}

%%%%%%%%%%%%%%%%%%%%%%%%%%%%%
\section{$\mathcal{N}=2$ Supersymmetry}
\label{s:N2gal}
%%%%%%%%%%%%%%%%%%%%%%%%%%%%%

For the purpose of deriving $\mathcal{N}=2$ SUSY Ward identities, it is very convenient to use superamplitudes. We use the super-wavefunction formalism \cite{Elvang:2011fx} (see also \cite{Elvang:2013cua,Elvang:2015rqa}) to encode the external states of the positive and negative helicity multiplets:
\be
  \label{superwave}
  \begin{split}
  \Phi^+ &= \gamma^+ + \eta_1 \lambda^{1+} + \eta_{2} \lambda^{2+} - \eta_1 \eta_2 Z\,,\\
  \Phi^- &= \,\bar{Z} \,+ \eta_1 \bar\lambda^{1-} + \eta_2 \bar\lambda^{2-} - \eta_1 \eta_2  \gamma^-\,.  
  \end{split}
\ee
The indices 1 and 2 are $SU(2)$ $R$-symmetry labels and $\eta_1,\eta_2$ are book-keeping Grassmann variables.  
The conjugate of $\lambda^{1+}$ is $\bar\lambda_1^{-}=-\bar\lambda^{2-}$ while the conjugate of $\lambda^{2+}$ is $\bar\lambda_2^{-}=\bar\lambda^{1-}$, since $SU(2)$ indices are lowered with $\eps_{21} = - \eps_{12} = 1$.

The complex scalar versions of the Galileon amplitudes $A_4(Z \bar{Z} Z \bar{Z})$ and $A_5(\bar{Z} Z \bar{Z}Z\bar{Z})$ are both in SUSY sectors with four powers of Mandelstams (two $\eta$'s for each $Z$ and none for $\bar{Z}$). This means that their superamplitudes are simply proportional to the Grassmann delta-function\footnote{This is the $\mathcal{N}=2$ equivalent of the MHV sector known from super-Yang-Mills theory or supergravity theories.}
\be
\delta^{(4)}_n(\tilde{Q}) = 
\frac{1}{2^2} \prod_{a=1,2} \sum_{i,j=1}^n \< i j\> \eta_{ia} \eta_{ja} \,.
\ee
We write the 4- and 5-point superamplitudes with the all-scalar amplitude as the basis amplitude \cite{Elvang:2009wd} as  
\bea
\label{superA4}
\mathcal{A}_4\big( \Phi^+ \Phi^- \Phi^+ \Phi^- \big)
&=& 
\frac{A_4(Z\bar{Z}Z\bar{Z})}{\<13\>^2}\,
\delta^{(4)}_4(\tilde{Q})\,,\\
\label{superA5}
\mathcal{A}_5\big( \Phi^- \Phi^+ \Phi^- \Phi^+ \Phi^-\big)
&=& \frac{A_5(\bar{Z}Z\bar{Z} Z \bar{Z})}{\<24\>^2}\,
\delta^{(4)}_5(\tilde{Q})\,.
\eea
One takes Grassmann-derivatives of these superamplitudes to project out component states according to \reef{superwave}. For example, with $\eta_{iA}$ being the Grassmann-variable of the $i^{\rm th}$ particle and $A=1,2$ the $SU(2)$ index, it follows from \reef{superwave}-\reef{superA4} that
\be
\label{N2SWI4pt}
A_4(\gamma^+ \gamma^- \gamma^+ \gamma^-)
=
\bigg(\frac{\partial}{\partial \eta_{21}}
\frac{\partial}{\partial \eta_{22}}\bigg)\,
\bigg(\frac{\partial}{\partial \eta_{41}}
\frac{\partial}{\partial \eta_{42}}\bigg)
\mathcal{A}_4\big( \Phi^+ \Phi^- \Phi^+ \Phi^- \big)
=
\frac{A_4(Z\bar{Z}Z\bar{Z})}{\<13\>^2}\<24\>^2\,.
\ee
This and similar $\mathcal{N}=2$ SUSY Ward identities are used in the following sections.

%%%%%%%%%%%%%%%%%%%%%%%%%%%%%
\subsection{No Quartic $\mathcal{N}=2$ Model with Two Real Galileons}
\label{s:N2SUSY4ptsigmaZ2}
%%%%%%%%%%%%%%%%%%%%%%%%%%%%%

It follows from \reef{N2SWI4pt} that the 4-photon amplitude is propertional to the 
4-scalar amplitude as 
\be
  \label{4ptSUSYWI1}
  A_4(Z \bar{Z} Z  \bar{Z}) = \frac{\<13\>^2}{\<24\>^2}
  A_4(\gamma^+ \gamma^- \gamma^+ \gamma^-) \,.
\ee
In an $\mathcal{N}=2$ model, the positive and negative helicity vector states are not related by supersymmetry, so  \reef{N4SUSYvec4pt} is not required to hold. 

As noted above \reef{4photon}, locality fixes the form of the 4-photon amplitude to be $A_4(\gamma^+ \gamma^- \gamma^+ \gamma^-) = \frac{1}{\Lambda^6}\<24\>^2[13]^2t$, hence \reef{4ptSUSYWI1} gives
\be
  \label{A4scalarN2}
  A_4(Z \bar{Z} Z  \bar{Z}) = \frac{1}{\Lambda^6}\<13\>^2[13]^2t 
  =  \frac{1}{\Lambda^6} t^3 \,.
\ee
As discussed around \reef{preA4ZZbZZb}-\reef{4ptsigmaZ1},  only the $s t u$ scalar amplitude is compatible with $\sigma_Z = 2$, not $t^3$; therefore we conclude that 
{\em an $\mathcal{N}=2$ quartic Galileon based on local 4-point interactions cannot have $\sigma_Z = 2$.}

%%%%%%%%%%%%%%%%%%%%%%%%%%%%%
\subsection{No Quartic $\mathcal{N}=2$ DBI-Galileon}
\label{s:N2Gal4sigma1}
%%%%%%%%%%%%%%%%%%%%%%%%%%%%%

Let us now consider $\sigma_Z=1$, i.e.~option (b) from Section \ref{s:review} with 
\be
 \label{sigmas}
  \sigma_\phi= 2\,,~~~
  \sigma_\chi= 1\,,~~~
  \sigma_\lambda= 1\,,~~~
  \sigma_\gamma= 0\,,
\ee
and the $\mathcal{N}=2$ compatible 4 scalar amplitude \reef{A4scalarN2}.  Projecting to the real scalar amplitudes using \reef{defZ} we find from \reef{A4scalarN2}  that 
\be
 A_4(\phi\phi\phi\phi) = A_4(\chi\chi\chi\chi)= \frac{3}{2\Lambda^6}\, stu\,,
\ee
while the mixed scalar amplitudes are
\be
 A_4(\phi\phi\chi\chi) = \frac{1}{2\Lambda^6} (t^3 - (s^3 + u^3))\,,~~~
 A_4(\phi\phi\phi\chi) =  A_4(\phi\chi\chi\chi) = 0\,.
\ee

To test the consistency of this model we use the soft bootstrap. At 6-point, the soft bootstrap is only valid for two amplitudes and they do not test the model beyond $\mathcal{N}=1$. (The details are presented in Appendix \ref{app:n4susy}.)  

To make progress, we therefore include DBI as the leading-order theory. In fact, not including DBI would not be in the spirit of EFT: all operators compatible with the symmetries should be included in the EFT and we have no symmetries to exclude DBI. Moreover, as discussed further in Section \ref{s:swampland}, UV-completability requires the presence of DBI. 

We consider three cases of quartic supersymmetric interactions:
\begin{itemize}
\item  DBI: In supersymmetric DBI, the 4-point amplitude of conjugate scalars $Z$ and $\bar{Z}$ takes the form $A_4(Z \bar{Z} Z  \bar{Z}) = \frac{1}{\Lambda^4}t^2$.
\item $\mathcal{N}=1$ Gal$_4$: The $\mathcal{N}=1$ compatible quartic Galileon with $A_4(Z \bar{Z} Z  \bar{Z}) = \frac{1}{\Lambda^6}s t u$ and $\sigma_Z=2$, and  
\item $\mathcal{N}=2$ Gal$_4$:
The candidate $\mathcal{N}=2$  quartic Galileon with $A_4(Z \bar{Z} Z  \bar{Z}) = \frac{1}{\Lambda^6} t^3$ and $\sigma_Z=1$.
\end{itemize}
It follows then from the superamplitude  \reef{superA4} that the 4-point amplitudes with two identical pairs of fermions take the form
\be
  A_4( \lambda^{1+} \bar\lambda ^{2-} \lambda^{1+} \bar\lambda ^{2-}) = - [13] \< 24\> f(s,t)\,,
\ee
with  
\be
  f^\text{DBI}(s,t) = t\,,~~~~
  f^\text{$\mathcal{N}=1$ Gal$_4$}(s,t) = su\,,~~~~
  f^\text{$\mathcal{N}=2$ Gal$_4$}(s,t) = t^2\,.
\ee
The difference between the two latter motivates testing the 6-fermion amplitude 
$A_6( \lambda^{1+} \bar\lambda ^{2-}  \lambda^{1+} \bar\lambda ^{2-}  \lambda^{1+} \bar\lambda ^{2-})$ with the soft bootstrap. For consistency, we consider three cases of the soft bootstrap where the 4-point vertices are 
\be
\label{3cases}
  \text{DBI} \times \text{DBI}\,,~~~~
  \text{DBI} \times \text{($\mathcal{N}=1$ Gal$_4$)}\,,
  ~~~\text{and}~~~
  \text{DBI} \times \text{($\mathcal{N}=2$ Gal$_4$)}\,.
\ee
The first two serve as test cases and they pass the $A_6( \lambda^{1+} \bar\lambda ^{2-}  \lambda^{1+} \bar\lambda ^{2-}  \lambda^{1+} \bar\lambda ^{2-})$ bootstrap test. The third fails the test. (For details, see Appendix \ref{app:n4susy}.) We also performed the bootstrap test  on $A_6( \lambda^{1+} \bar\lambda ^{2-}  \lambda^{1+} \bar\lambda ^{2-}  \phi\phi)$ with the same outcome: the first two case in \reef{3cases} pass but 
DBI $\times$ ($\mathcal{N}=2$ Gal$_4$) does not. 
This shows that {\em the quartic DBI-Galileon is not compatible with $\mathcal{N}=2$ supersymmetry}.

Without DBI, the $\mathcal{N}=2$ quartic Galileon remains a candidate since there are no non-trivial soft bootstrap checks that can be used to test its consistency. However, it is in the Swampland, as discussed further in Section \ref{s:swampland}.

The soft bootstrap analysis assumed $\sigma_\lambda =1$ for the fermions. 
One could entertain the idea of a supersymmetric Galileon in which the fermions had $\sigma_\lambda=0$ instead of $\sigma_\lambda=1$, however this gets into the territory of very few constraints on the model and it hardly falls in the category of an ``exceptional" EFT.

%%%%%%%%%%%%%%%%%%%%%%%%%%%%%
\subsection{Quintic $\mathcal{N}=2$ Galileon-Axion Model}
\label{s:N2quinticsigmaZ1}
%%%%%%%%%%%%%%%%%%%%%%%%%%%%%

From the superamplitude \reef{superA5} we can obtain the $\mathcal{N}=2$ SUSY Ward identity\footnote{This is also the conjugate of the SUSY Ward identity 
\reef{fSUSY1} which was an intermediate $\mathcal{N}=2$ result on the way to the $\mathcal{N}=4$ SUSY Ward identity \reef{5ZN4}.}
\be
  \label{A5SUSYWI1}
  A_5^\text{$\mathcal{N}=2$ Gal$_5$}(\gamma^-\gamma^+\gamma^-\gamma^+\bar{Z})
  =
  -\frac{\<13\>^2}{\<24\>^2}  
  \, A_5^\text{$\mathcal{N}=2$ Gal$_5$}(\bar{Z}Z\bar{Z} Z \bar{Z})\,.
\ee 

The amplitude on the RHS is the familiar 5-scalar amplitude for which the 3-parameter family \reef{A5scalar3par} is compatible with $\mathcal{N}=1$ SUSY. 
There are 10 independent possible local matrix elements for the amplitude on the LHS of \reef{A5SUSYWI1}. This reduces to 8 parameters after imposing $\sigma_Z =1$; no further constraints arise from requiring $\sigma_\phi =2$. The Ward identity \reef{A5SUSYWI1} with \reef{A5scalar3par} on the RHS is solved for 
\be  
 \label{N2susycond}
 \mathcal{N}=2 ~ \text{Gal$_5$:} ~~~~ q_3 = 3 q_2\,,
\ee 
with all 8 parameters on the LHS fixed in terms of $q_1$ and $q_2$. 
 There are no further constraints from $\mathcal{N}=2$ Ward identities at 5-point. Thus we find that a 2-parameter family of solutions for the 5-point Galileon local matrix elements compatible with $\mathcal{N}=2$ supersymmetry.

We have performed tests with the soft bootstrap of all constructible 7-point amplitudes, as outlined in Appendix \ref{app:n5}, using one $\mathcal{N}=2$ DBI 4-point vertex and one  5-point vertex from the family of the $\mathcal{N}=2$ quintic Galileons. We have not found any restrictions on the two parameters $q_1$ and $q_2$. Therefore we cannot exclude the existence of a quintic Galileon compatible with $\mathcal{N}=2$ supersymmetry.

%%%%%%%%%%%%%%%%%%%%%%%%%%%%%
\section{Cubic Interactions and the Emergence of Gravity}
\label{s:SG}
%%%%%%%%%%%%%%%%%%%%%%%%%%%%%

Our analysis of extended supersymmetry in Sections \ref{s:noN4Gal} and \ref{s:N2gal}  assumed that the 4- and 5-point matrix elements, connected to the quartic and quintic Galileons respectively, are all from local operators. In other words, the 4- and 5-point matrix elements were assumed not to have poles, which is justified by the absence of 3-point interactions. In this section, we relax that assumption.  

Let us examine what happens if we insist on $\sigma_Z=2$, but allow $A_4(\gamma^+ \gamma^- \gamma^+ \gamma^-)$ to have poles. To be compatible with softness $\sigma_Z = 2$ we must have 
$A_4(Z \bar{Z} Z  \bar{Z}) = s t u/\Lambda^6$. It then follows from the SUSY Ward identity \reef{4ptSUSYWI1} that 
\be
  \label{4ptSUSYWI1nonloc}
  A_4(\gamma^+ \gamma^- \gamma^+ \gamma^-)
  =\frac{\<24\>^2}{\<13\>^2}
  A_4(Z \bar{Z} Z  \bar{Z})
  =
  \frac{1}{\Lambda^6}\frac{[13]^2\<24\>^2s u}{t}\,,
\ee
so there is a pole in the $t$-channel. On this pole, which has  $P^2=(p_1+p_3)^2 = 0$, the amplitude  factorizes into the 3-particle amplitudes 
\be
  \label{A3wgrav}
  A_3 (\gamma_1^+ \gamma_3^+ h_{-P}^+) = \frac{1}{\Lambda^3}[1P]^2[3P]^2\,,~~~~
  A_3 (\gamma_2^- \gamma_4^- h_P^-) = \frac{1}{\Lambda^3}\<2P\>^2\<4P\>^2\,.~~~~
\ee
The particle exchanged is a massless spin-2 particle, i.e.~a graviton. The matrix elements \reef{A3wgrav} arise from the cubic interaction of the operator $R_{\mu\nu} F^{\mu\lambda} F^{\nu}_{~\lambda}$. So, apparently one needs a graviton in the spectrum in order to have a pair of Galileons in the scalar components of an interacting $\mathcal{N}=2$ vector multiplet!

Then comes the question if the Galileons and the graviton are part of the same supermultiplet or not. Let us briefly discuss some options:\footnote{We are grateful for discussions of some of these cases with Shruti Paranjape and Callum Jones.} 
\begin{itemize}
\item $\mathcal{N}=4$ supersymmetry I: Galileons and gravitons in {\em separate} supermultiplets.

In this case, the Galileons must be part of a massless $\mathcal{N}=4$  CTP self-conjugate vector supermultiplet and the SUSY Ward identity \reef{N4SUSYvec4pt} has to hold. However, the 4-vector amplitude \reef{4ptSUSYWI1nonloc} is incompatible with  \reef{N4SUSYvec4pt}, hence this option is  excluded.
\item $\mathcal{N}=4$ supersymmetry II: Galileons and gravitons in the {\em same} supermultiplet.

The simplest option is to consider the Galileons to be the scalar component of the $\mathcal{N}=4$ supergravity multiplet. In that case, the Galileon $\mathcal{N}=2$ on-shell superfield \reef{superwave}
is enhanced to an $\mathcal{N}=4$ on-shell superfield whose top component is the graviton:
\be
  \label{superwaveN4}
  \begin{split}
  \mathcal{H}^+ &= h^+ + \eta_A \psi^{A+} \,
  + \eta_A\eta_B \gamma^{AB+} 
  + \eta_A\eta_B\eta_C\lambda^{ABC+} 
  + \eta_1\eta_2 \eta_3\eta_4 Z\,,\\
  \mathcal{H}^- &= \,\bar{Z} \,
  + \eta_A \bar\lambda^{A-} + \eta_A \eta_B  \gamma^{AB-}
  + \eta_A\eta_B\eta_C \psi^{ABC-}
  + \eta_1\eta_2 \eta_3\eta_4  h^-
  \,,
  \end{split}
\ee
where $\psi$ are the gravitinos and $A,B,C,D=1,2,3,4$. 
Supersymmetry then relates a (higher-derivative) 4-graviton amplitude to the Galileon 4-scalar amplitude via  
\be
  \label{grav2Z}
  A_4(h^+ h^- h^+ h^-)
  = \frac{\<24\>^4}{\<13\>^4}
  A_4(Z \bar{Z} Z  \bar{Z})\,.
\ee
For  $\sigma_Z=2$, this then gives
\be
 A_4(h^+ h^- h^+ h^-) = \frac{\<24\>^4 [13] s u}{\<13\>^3}\,
 ~~~\text{(!)}
\ee
This expression has a 3rd order pole and it is therefore not a physical amplitude. Hence the model is incompatible with locality: no such local field theory exists. 
\item $\mathcal{N}=2$ supersymmetry with $\sigma_Z=2$:\\[1mm]
This case involves an $\mathcal{N}=2$ matter vector multiplet containing two Galileons in the complex scalar $Z$ with $\sigma_Z=2$. There graviton must be part of a separate $\mathcal{N}=2$ 
supergravity multiplet.

This option has a curious outcome. Consider first the 3-particle interactions \reef{A3wgrav}. All three states are top-components of their respective $\mathcal{N}=2$ supermultiplets. Thus if $Q$ raises helicity by $1/2$, it will imply the vanishing of $A_3 (\gamma^+ \gamma^+ h^+)$. The only way around this is if the  graviton is {\em not} the top-component of its supermultiplet: then instead $A_3 (\gamma^+ \gamma^+ h^+)$ becomes related to an amplitude $A_3 (\lambda^+ \gamma^+ X^+)$, where $X$ a massless spin 5/2 particle. This sounds peculiar, but it is also implied directly from 4-point as we shall see now.

Together with $A_4(Z \bar{Z} Z  \bar{Z}) = s t u/\Lambda^6$, the  4-point $\mathcal{N}=2$ superamplitude \reef{superA4} with superwavefunctions \reef{superwave} gives
\be
  \label{weirdones}
  A_4(\gamma^+ \gamma^- \lambda^{1+} \lambda^{2-})
  = -\frac{\<23\>\<24\> [13]^2 s u}{t}\,,
  ~~~~~
  A_4(\gamma^+ \gamma^- Z \bar{Z})
  = \frac{\<23\>^2 [13]^2 s u}{t}\,.\\
\ee
Both have $t$-channel poles and in the first case a massless spin 5/2 particle is exchanged via 3-point interactions precisely like those described above. The second amplitude in \reef{weirdones} has a spin 3 massless particle exchanged. This means that the graviton must be a part of a supermultiplet with a spin 3 and two spin 5/2 particles. 

This is of course rather bizarre and the immediate intuition may be that these models with their massless higher-spin states should be junked. This is based on the well-known results that higher-spin massless particles cannot couple consistently to particles with spin $< 2$ in flat Minkowski space. Such results can be derived from the soft behavior of amplitudes, however, this assumes  ``normal'' gravitational interactions. In particular, in the case here,  the regular 2-derivative gravitational interactions with $[\kappa]=-1$ are assumed to be absent (removed by some suitable scaling limit) and instead we have the higher-derivative 3-point interactions \reef{A3wgrav} with coupling dimension $[g] = -3$.

So, from the low-energy point of view, this higher-derivative model with two real Galileons in an $\mathcal{N}=2$ vector supermultiplet  coupled to a spin 
$(2,\,5/2,\, 3)$  $\mathcal{N}=2$ supergravity multiplet appears viable. However, these models can be excluded because such higher-derivative higher-spin operators lead to asymptotic time-advance  \cite{Hinterbichler:2017qcl}. So, this model is excluded too.
\end{itemize}

The above discussion was motivated by the graviton exchange in the vector amplitude \reef{4ptSUSYWI1nonloc} for the $\sigma_Z=2$ case. Yet, we may also entertain the idea of a model with the Galileon $\phi$ and the $R$-axion $\chi$
 as the scalars in the $\mathcal{N}=4$ supergravity multiplet. Now we take $A_4(Z \bar{Z} Z  \bar{Z}) = t^3/\Lambda^6$ and supersymmetry then gives the local result \reef{4photon} for  the 4-vector amplitude is polynomial. We know this cannot be coupled consistently with DBI (Section \ref{s:N2Gal4sigma1}), but suppose we study it in its own right. 
\begin{itemize}
\item $\mathcal{N}=4$ supergravity with one  Galileon and one $R$-axion ($\sigma_Z=1$).\\[1mm]
Let the Galileon and $R$-axion be the scalar component of an $\mathcal{N}=4$ supergravity multiplet \reef{superwaveN4}. 
The 4-graviton amplitude computed from \reef{grav2Z} is now  
\be
  A_4(h^+ h^- h^+ h^-) = \frac{\<24\>^4 [13]^4}{t}\,.
\ee
It has a simple physical pole corresponding to a scalar exchange. The associated 3-point amplitudes come from 3-particle interactions 
\be
 \label{RRZ}
Z R_- R_- + \bar{Z} R_+ R_+\,,
\ee 
where the $\pm$ subscripts on the Ricci's refer to the 4d spinorized form of curvature tensors which are in one-to-one correspondence with the graviton helicity states. These interactions are naively compatible with supersymmetry in a standard sense. 
\item $\mathcal{N}=2$ supergravity + $\mathcal{N}=2$ matter.\\[1mm]
Consider the case of $\mathcal{N}=2$ supergravity coupled with an $\mathcal{N}=2$ matter vector multiplet with $\sigma_Z=1$. In this case, there are no poles in the 4-point Galileon superamplitudes, hence there is no clear link between the two supermultiplets. 
\end{itemize}

In the scenarios outlined right above, the 3-particle interactions may jeopardize the soft behavior of the scalars.\footnote{See for example \cite{Elvang:2018dco} where it was found that the $\mathcal{N}=2$ $\mathbb{CP}^1$ model has vector-scalar amplitudes with non-vanishing soft scalar limits.}  To assess this, one has to compute higher-point amplitudes and examine whether the vanishing soft scalar limits with weights $\sigma_\phi=2$ or $\sigma_\chi =1$ continue to hold. This is beyond the scope of this paper and we leave it for future work to investigate these possibilities further. Any connection to some decoupling limit of massive gravity might be of interest.

%%%%%%%%%%%%%%%%%%%%%%%%%%%%%
\section{UV-Completability vs.~the Swampland}
\label{s:swampland}
%%%%%%%%%%%%%%%%%%%%%%%%%%%%%

In this section, we discuss when the quartic Galileon can appear in low-energy effective actions of some UV complete theory. We further discuss the recently found numerical constraints on the quartic Galileon coupling the DBI-Galileon. 

\subsection{Only the DBI-Galileon is UV-Completable}

Consider the low-energy effective action of a single real scalar $\phi$, such as
\be
  \label{genphiL}
  \mathcal{L} = -\frac{1}{2}(\partial \phi)^2
  + \frac{c_4}{\Lambda^4} (\partial \phi)^4 
  - \frac{c_6}{\Lambda^6}
  (\partial \phi)^2(\partial \partial \phi)^2
  + \ldots
\ee
In a generic bottom-up setting, there are no constraints on the value of the dimensionless numbers $c_4$, $c_6$, \ldots. However, if the model is to arise from sensible UV physics, i.e.~be a ``UV-completable model", then is must be that 
\be 
  \label{c4}
  c_4 > 0\,.
\ee  
This follows from a sum rule based on the optical theorem which relates the imaginary value of  the 4-particle amplitude in the forward limit $t \to 0$ to the total  cross-section $\sigma(s)$. Since $\sigma(s)>0$, the integral
\be
   \label{sumrule}
   \int \frac{\text{Im}(A_4(s,0))\,ds}{s^{3}} =
   \int \frac{\sigma(s)\,ds}{s^{2}} > 0
\ee
must be strictly positive \cite{Adams:2006sv,Arkani-Hamed:2020blm}. The 4-scalar amplitude of \reef{genphiL} is 
\be
  \label{4ptDBIGal4}
  A_4(s,t) = 
  \frac{2c_4}{\Lambda^4}\big(s^2 + t^2 +u^2\big)
  + \frac{3 c_6}{\Lambda^6} s t u + \ldots \,.
\ee
In the forward limit $t \to 0$, we see that $A_4(s,0)/s^3$ has a simple pole at $s=0$ with 
residue  $4c_4/\Lambda^4$. A contour deformation of the integral on the LHS of \reef{sumrule}  picks up this residue and then it follows that $c_4$ must be positive, as stated in \reef{c4}.

The sum rule argument assumes convergence of the dispersion integral \reef{sumrule} and that the contour deformation described above can be performed  without picking up contributions from the contour-segments at infinity. This requires that the forward amplitude cannot grow faster than $|s|^2$ at large $|s|$. That in turn is guaranteed by the Froissart bound \cite{Froissart:1961ux,Martin:1962rt}, which states that $\sigma(s) < \ln^2|s|$ at large $|s|$. This is   violated only if the UV theory has highly unusual behavior \cite{Adams:2006sv}. Therefore, with the usual assumptions of locality, unitarity, and analyticity, the positivity bound \reef{c4} is a necessary condition for the existence of a sensible UV theory, i.e.~that the model is  ``UV-completability". An effective field theory which does not satisfy the UV-completability criteria can be said to be in the Swampland.

\noindent {\bf Example: DBI.} 
The DBI Lagrangian realizes the positivity constraints \reef{c4}: the low-energy expansion of \reef{DBI} can be written as
\be
  \label{DBIexpanded}
  \begin{split}
  \mathcal{L}_\text{DBI}
   &= - \Lambda^4 
  \bigg( 
  \sqrt{1 - \tfrac{1}{\Lambda^4}(\partial \phi)^2}
  -1
  \bigg)\\
  &=
  -\frac{1}{2} (\partial \phi)^2
  + \frac{1}{8 \Lambda^4} (\partial \phi)^4
  - \frac{1}{16 \Lambda^8} (\partial \phi)^6
  + \frac{5}{128 \Lambda^{12}} (\partial \phi)^8
  + \ldots
  \,
  \end{split}
\ee
and clearly $c_4 = \frac{1}{8} >0$.

\noindent {\bf Example: Quartic Galileon.} Consider now the quartic Galileon \reef{gal4action} with $g_4/3 = c_6/\Lambda^6$:
\be
  \label{badgal4}
  \mathcal{L} = -\frac{1}{2}(\partial \phi)^2
  +
  \frac{c_6}{\Lambda^6} (\partial \phi)^2
  \big(
     (\Box \phi)^2 - (\partial\partial\phi)^2
  \big) + \ldots\, , 
\ee
The term with $\Box \phi$ does not contribute to the on-shell 4-point amplitude, so the 4-point amplitude is \reef{4ptDBIGal4} with $c_4=0$. This model cannot arise as the low-energy limit of a theory with sensible UV physics since it violates the sum rule positivity constraint \reef{c4}. This means that the quartic Galileon is a Swampland model.

\noindent {\bf Example: Quartic DBI-Galileon.}
The Galileon can avoid the Swampland when it is  ``protected" by a leading order $(\partial \phi)^4$ term with a positive coefficient $c_4 > 0$ in order to satisfy the sum rule \reef{sumrule} for $k=1$.

There is only one way the $(\partial \phi)^4$ term can realize the enhanced Galileon shift-symmetry \reef{shiftsym}, and that is as part of DBI. So the quartic DBI-Galileon does satisfy this basic  UV-completability criterion. 
 
%%%%%%%%%%%%%
\subsection{Constraints on the DBI-Galileon Couplings}

Consider the DBI-Galileons
\be
  \label{DBIgal4}
  \mathcal{L}_\text{DBI-Gal4}
  = -\frac{1}{2}(\partial \phi)^2
   +\frac{c_4}{\Lambda^4}(\partial \phi)^2
    +
  \frac{c_6}{\Lambda^6} (\partial \phi)^2
  \big(
     (\Box \phi)^2 - (\partial\partial\phi)^2
  \big)
  + \ldots\, ,
\ee
with ``$+\ldots$" denoting both higher-point and  higher-derivative terms. The 4-point amplitude is \reef{4ptDBIGal4}.

In a recent work \cite{Caron-Huot:2020cmc,Bellazzini:2020cot,Tolley:2020gtv,Guerrieri:2020bto}, further constraints from UV-completability were derived, in particular bounds on $c_6$ in terms of $c_4$. Translating the bounds in \cite{Caron-Huot:2020cmc} to the parameterization above, we find that
\be
  -3.449 < \frac{c_6}{c_4} \le 1\,.
\ee
When we take $c_4 = 1/8$, as in \reef{DBIexpanded} this gives a small allowed range for the Galileon quartic coupling, namely
\be
  \label{simonbnd}
  -0.431 < c_6 \le \frac{1}{8}= 0.125 \,.
\ee
 
In this paper, we have shown that quartic DBI-Galileons are incompatible with  $\mathcal{N}>1$  supersymmetry.\footnote{In the absence of cubic interactions.} The absence of the Galileon, i.e.~$c_6=0$, is compatible with the bound \reef{simonbnd}, as it should be since the superstring amplitude has $c_6=0$.

In what contexts may quartic DBI-Galileons then show up? For the bosonic string, it is tricky to apply the positivity constraints because of the presence of the tachyon in the spectrum. We have ruled out $\mathcal{N}>1$  supersymmetry. The leaves potentially  configurations of branes preserving $\mathcal{N}=1$ supersymmetry. It would be interesting to perform an S-matrix bootstrap analysis such as \cite{Caron-Huot:2020cmc,Bellazzini:2020cot,Tolley:2020gtv,Guerrieri:2020bto} with the additional assumption of $\mathcal{N}=1$ supersymmetry to understand if that would further constrain the Galileon coupling. 

%%%%%%%%%%%%%%%%%%%%%%%%%%%%%
%%%%%%%%%%%%%%%%%%%%%%%%%%%%%
%%%%%%%%%%%%%%%%%%%%%%%%%%%%%
\section{Discussion}
\label{s:discussion}
%%%%%%%%%%%%%%%%%%%%%%%%%%%%%
%%%%%%%%%%%%%%%%%%%%%%%%%%%%%
%%%%%%%%%%%%%%%%%%%%%%%%%%%%%
In this paper we have taken a bottom-up EFT  approach to study the possibilities for supersymmetrization of Galileons and DBI-Galileons. The results were summarized in the Introduction. Let us now discuss our results in perspective of the effective 3-brane construction of these models. 

As shown in \cite{deRham:2010eu}, the single-scalar 4d DBI-Galileon can be derived from
\begin{equation} \label{DBIGal}
  S = \int \text{d}^4x \sqrt{-G} \Big[\Lambda^4_2 + \Lambda^3_3 K[G]+\Lambda^2_4 R[G]+\Lambda_5 \mathcal{K}_{\text{GHY}}[G]\Big]\,,
\end{equation}
where $G$ is the pullback of the 5d bulk metric to the 3-brane. DBI arises from  the term $\Lambda_2^4$ and the subleading terms give the  Galileon interactions: the extrinsic curvature $R[G]$ is responsible for the quartic Galileon while the cubic and quintic Galileons stem from the intrinsic curvature   $K[G]$ and the Gibbons-Hawking-York (GHY) $\mathcal{K}_{\text{GHY}}[G]$. The latter two are possible only as boundary terms for an end-of-the-world brane. Let us first discuss the quintic Galileon and then the quartic.

\noindent {\bf Non-SUSY Quintic Galileons with $\sigma_Z =2$.}
A 4d model with two Galileon scalars  naturally arises from a 3-brane with two transverse directions. The spacetime rotation among those two directions induces an $SO(2)$ rotation symmetry of the two real scalars on the brane. This symmetry cannot be preserved by odd-point scalar amplitudes, so this raises the question of the physical interpretation of the non-supersymmetric quintic complex scalar model \reef{A5sigmaZ2} with $\sigma_Z =2$. To examine this, project the real scalar 5-point amplitudes from 
$A_5(Z \bar{Z} Z \bar{Z} Z) =  g_5' \,\big(\eps(1,2,3,4)\big)^2$ using \reef{defZ}. One finds that
\be
  \label{A5sreal1}
  A_5(\phi\phi\phi\phi\phi)\,,~~~~
  A_5(\phi\phi\phi\chi\chi)\,,~~~~
  A_5(\phi\chi\chi\chi\chi)\,
\ee
are all non-zero and proportional to $(\eps(1,2,3,4))^2$ while
\be
  \label{A5sreal2}
  A_5(\phi\phi\phi\phi\chi)=
  A_5(\phi\phi\chi\chi\chi)=
  A_5(\chi\chi\chi\chi\chi)= 0\,.
\ee
Thus, in this model, only $\phi$ has a local quintic self-interaction. Soft recursion ensures that the all-$\phi$ tree amplitudes are precisely those of the real-scalar quintic Galileon because $A_5(\phi\phi\phi\phi\chi)=0$ ensures that there can be no internal $\chi$-exchanges. The scalar $\chi$ has no local quintic self-interaction and it only interacts via vertices that involve $\phi$. Moreover, the vanishing of the amplitudes \reef{A5sreal2} indicates that $\chi$ has a symmetry $\chi \to -\chi$. It is clear that the two scalars $\phi$ and $\chi$ are not on the same footing and that there is no exchange symmetry, let alone $SO(2)$ symmetry, between them. 

Here is a possible physical picture of this model. Consider the low-energy EFT on a 3-brane in 6d flat space with one transverse direction that extends on both sides of the brane, giving rise to $\chi$, and a second transverse direction for which the brane is an end-of-the-world brane, giving rise to $\phi$. The inequivalence of the two transverse directions breaks the $SO(2)$ symmetry and allows the scalar 5-point  amplitudes \reef{A5sreal1} to be non-zero. In particular, the quintic interactions could be generated by the GHY term in \reef{DBIGal}. 
This model can additionally have the usual DBI interactions for both $\phi$ and $\chi$ as well as the quartic Galileon interactions when 
 $\Lambda_2$ and $\Lambda_4$ are non-zero in \reef{DBIGal}. It would be interesting to carry out this construction in detail and see if it matches the amplitudes \reef{A5sreal1}  and \reef{A5sreal2}.

The two-Galileon ($\sigma_Z =2$) quintic model \reef{A5sigmaZ2} was shown in  \cite{Elvang:2017mdq} to be incompatible with $\mathcal{N}=1$ supersymmetry. This assumes  that all 5-point interactions are local (i.e.~polynomial and free of any poles), just as discussed above. We have also seen in Section \ref{s:SG} that poles in the quartic amplitudes may in some cases be interpreted in a supergravity context and it would be interesting to examine if the same may be the case for the $\sigma_Z =2$ quintic model \reef{A5sigmaZ2}. We leave this for future investigation.

\noindent {\bf SUSY Quintic Galileons with $\mathbf{\sigma_Z =1}$.}
The supersymmetrizations of the quintic Galileon with only local 5-point interactions have the Galileon $\phi$ coupled with an $R$-axion $\chi$. This model presumably arises from an end-of-the-world 3-brane in 5d flat space. The scalar $\phi$ is the Goldstone mode of broken transverse direction while $\chi$ arises from the breaking of an $R$-symmetry. The fermions are Goldstinos of broken SUSY generators. Having a more precise picture of the $\mathcal{N}=1$ and $\mathcal{N}=2$ supersymmetric models with their respective 3 and 2 free parameters would be interesting from the point of view of a generalization of brane embeddings such as \reef{DBIGal}. In particular, it would be interesting to understand how more than one quintic interaction can arise in multi-scalar models from brane constructions like \reef{DBIGal} --- or perhaps it could be used to restrict the models further. 
Likewise it would be useful to understand in this picture why $\mathcal{N}=4$ (with spins $\le 1$) is disallowed --- or whether it could be allowed when coupled to supergravity. 

A number of papers have constructed models with two or more Galileons, see for example \cite{Padilla:2010de,deFromont:2013iwa,Allys:2016hfl} for Lagrangian-based approaches and \cite{Kampf:2020tne} for amplitude and soft limit constructions. Often $SO(N)$ or $SU(N)$ symmetry is assumed. Similarly \cite{Kampf:2021bet} considers Galileon-vector models. It could be good to understand their connection to the on-shell amplitude constructions considered in this paper and in \cite{Elvang:2017mdq,Elvang:2018dco}.

\noindent {\bf Quartic Galileons.}
Since the quartic Galileon arises from the Ricci-term in \reef{DBIGal}, one might have  expected it to be compatible with extended supersymmetry. So it is perhaps surprising that we find it to be so heavily constrained for $\mathcal{N}>1$. Recall that in absence of gravity, we have shown that an $\mathcal{N}=4$ quartic Galileon is impossible while the  $\mathcal{N}=2$ vector multiplet is allowed only when only for a single Galileon joined by an $R$-axion (and not another Galileon) and in that case only in the absence of DBI. I.e.~the quartic DBI-Galileon is not compatible with $\mathcal{N}=2$ SUSY. This leaves the only option for $\mathcal{N}=2$ quartic Galileons in the Swampland, as discussed in Section \ref{s:swampland}. (We  discussed options with gravity in Section \ref{s:SG}.) Understanding these amplitude-based results from the point of view of the brane embeddings \reef{DBIGal} would be useful.

\noindent {\bf Massive Gravity.}
Galileons also occur in limits of massive gravity. Understanding (limits on) supersymmetrization of massive gravity may shed light on our results for supersymmetrizations of Galileons and perhaps clarify why in some cases supersymmetry appears to lead us to coupling between Galileons and gravitons. If these arise from limits of supersymmetric massive gravity it would appear that the extended supersymmetry constraints no longer permits the decoupling of Galileons from gravity. 

\noindent {\bf Hypermultiplets.} 
Let us finally note that a supersymmetrization we have not considered in this paper is the option of Galileons (possibly joined by $R$-axions) as part of $\mathcal{N}=2$ hypermultiplets. Future studies may shed light on such constructions. 

\noindent {\bf Double-Copy.} 
Let us now discuss DBI and Galileons in the context of the double-copy.
DBI arises from the double-copy as 
\be
  \label{N4sDBIdc}
  (\mathcal{N} = 4~\text{SUSY DBI}) = (\mathcal{N} = 4~\text{SYM})  \otimes \chi\text{PT} \,.
\ee
The single-scalar quartic Galileon can be obtained as a double-copy from
\be
  \label{gal4dc}
    \text{Gal}_4 =   \chi\text{PT}\otimes \chi\text{PT} \,.
\ee
Since $\chi\text{PT}$ is not supersymmetrizable, no supersymmetric version of the Galileon can be obtained from a simple supersymmetrization of \reef{gal4dc}. 

Because Galileons can be thought of as subleading terms to DBI, so one might have  hoped to generate supersymmetric quartic Galileons from \reef{N4sDBIdc} with higher-derivative corrections (h.d.), such as
\be
  \label{hdN4sDBIdc}
  (\mathcal{N} = 4~\text{SUSY DBI}+ \text{h.d.}) = (\mathcal{N} = 4~\text{SYM}+ \text{h.d.})  \KLT{gen.} ( \chi\text{PT}+ \text{h.d.})\,,
\ee
where the double-copy may be done with the generalized KLT kernel found recently via the double-copy bootstrap \cite{Chi:2021mio}. 
However, this fails to produce any $O(p^6)$ terms at 4-point. The results in this paper explain why: we have shown that there is no $\mathcal{N} = 4$ supersymmetrization of the quartic Galileon (in the absence of gravity), so indeed the double-copy \reef{hdN4sDBIdc} cannot produce it. 

Next, consider
\be
  \label{hdN4sDBIdc2}
  (\mathcal{N} = 2~\text{SUSY DBI}+ \text{h.d.}) = (\mathcal{N} = 2~\text{SYM}+ \text{h.d.})  \KLT{h.d.} ( \chi\text{PT}+ \text{h.d.})\,.
\ee
Again, this does not produce $\mathcal{N} = 2$ quartic Galileons, and we understand why: the quartic DBI-Galileon is not compatible with $\mathcal{N} = 2$ supersymmetry.

The $\mathcal{N}=1$ and non-susy versions of \reef{hdN4sDBIdc2} do not produce any scalars in the double-copy: 
$\text{YM}\otimes\chi\text{PT}$ simply gives the Born-Infeld (BI) model  with a spectrum of only the abelian photons. One cannot just decouple the gluons from SYM and try such a double-copy with $\chi\text{PT}$ because the SYM scalar tree amplitudes include gluon exchanges. So we have two puzzles:
\begin{itemize}
\item why does (any known consistent versions of) the double-copy not produce Galileons as subleading terms to DBI? 
\item why does the double-copy appear to be unable to produce a supersymmetric version of the Galileons? 
\end{itemize}
There is similar a puzzle for the quintic Galileon which is not produced in any known double-copy \cite{Elvang:2018dco}. 
Perhaps future work on the EFTs, brane-constructions, and the double-copy will shed light on these questions.

\noindent {\bf String Amplitudes.}
In the open superstring, we know Galileon terms are absent and we now understand this from the point of view of compatibility with supersymmetry. The bosonic string with external abelian vectors has a low-energy effective action that includes $O(p^6)$ terms of the form $\partial^2 F^4$ \cite{Tseytlin:1987ww,Andreev:1988cb,Tseytlin:1999dj}. In the absence of SUSY, they cannot be linked to $O(p^6)$ terms in the scalar bosonic string effective action, but a direct calculation\footnote{We thank Stephan Stieberger for communications about this point.} shows that they do indeed occur.\\[2mm]

Together with \cite{Elvang:2017mdq,Elvang:2018dco} the results of this paper shows that the possibilities for supersymmetry of 4d Galileons is highly constrained with only a few cases passing the bottom-up EFT constraints. Understanding this from other directions, such as brane constructions and massive gravity, is an interesting future direction.

%%%%%%%%%%%%%%%%%%%%%%%%%%%%%%
\section*{Acknowledgements}
%%%%%%%%%%%%%%%%%%%%%%%%%%%%%%
We would like to thank Clifford Cheung, Marios Hadjiantonis, Aidan Herderschee, Kurt Hinterbichler, Yu-tin Huang, Callum Jones,  Shruti Paranjape, and Stephan Stieberger for comments and great discussions. This work was in part supported by Department of Energy grant DE-SC0007859.  

%%%%%%%%%%%%%%%%%%%%%%%%%%%%%%
\appendix
%%%%%%%%%%%%%%%%%%%%%%%%%%%%%%

%%%%%%%%%%%%%%%%%%%%%%%%%%%%%%

\appendix

%%%%%%%%%%%%%%%%%%%%%%%%%%%%%
\section{Review of the Soft Bootstrap}
\label{app:softboot}
%%%%%%%%%%%%%%%%%%%%%%%%%%%%%

The soft bootstrap is based on the {\em soft subtracted recursion relations}  developed in \cite{Cheung:2014dqa,Kampf:2013vha,Cheung:2015cba,Cheung:2015ota,Cheung:2016drk} for scalar theories. In this appendix we briefly review first these recursion relations and the validity criteria derived in \cite{Elvang:2018dco}, then we explain how this is used to bootstrap the space of exceptional effective field theories for which the soft behavior of massless states is non-trivial. Finally we provide some details of how the soft bootstrap is used in the maintext of the paper. 

%%%%%%%%%%%%%%%%%%%%%%%%%
\subsection{Soft Subtracted Recursion and Its Validity Criterion}
%%%%%%%%%%%%%%%%%%%%%%%%%

 The idea of the soft subtracted recursion relations is to exploit the soft behavior of amplitudes \reef{softAn} to improve the large-$z$ behavior of amplitudes  under shifts
\be
  \label{phat}
  \hat{p}_i = (1-a_i z) p_i \,,
  ~~~\text{assuming}~~~~
  \sum_{i=1}^n a_i p_i = 0\,.
\ee
The latter ensures $n$-point momentum conservation of the shifted momenta.  
We assume all $a_i$ to be distinct such that the limit $z \to 1/a_i$  is the  single soft limit implemented as in \reef{softconv} with $\eps=(1-a_i z)$.

The subtracted recursion relations are derived from the contour integral 
\be
   \label{oint}
   \oint \frac{dz}{z} \frac{\hat{A}_n(z)}{F(z)} 
   ~~~\text{with}~~~
   F(z) = \prod_{i=1}^n (1-a_i z)^{\sigma_i} \,.
\ee
Here the deformed amplitude $\hat{A}_n(z)$ is the on-shell amplitude as a function of the shift momenta \reef{phat}. The $\sigma_i$ is the soft weight \reef{softAn} of the $i^\text{th}$ particle, so $\hat{A}_n(z)/F(z)$  does not have any poles at $z=1/a_i$. Therefore the only poles of the integrand \reef{oint} are those of the deformed amplitude $\hat{A}_n(z)$ corresponding to exchanges of physical particles and the pole at $z=0$. The residue of the latter is exactly the undeformed amplitude $A_n = \hat{A}_n(z=0)$.

In the absence of a simple pole at $z= \infty$, the integral \reef{oint} vanishes when the contour is chosen to surround all the poles at finite $z$. Thus, the residue of the pole at $z=0$ is equal to minus the sum of residues at $z \ne 0$. This means that the undeformed amplitude can be written as
\be
  \label{softsubrecrel1}
  A_n = \hat{A}_n(z=0) =  \sum_{I,  h_I,\pm} 
  \frac{\hat{A}^\text{L}_n(z_I^\pm)\hat{A}^\text{R}_n(z_I^\pm)}{F(z_I^\pm) P_I^2 (1-z_I^\pm / z_I^\mp)} \,,
\ee
where the sum is over all factorization channels $I$, spectrum of particles exchanged in the channel $I$, and the two solutions $z_I^\pm$ to the quadratic on-shell condition $\hat{P}_I^2 = 0$.

The form \reef{softsubrecrel1} of the soft subtracted recursion relations is somewhat complicated to work with due to the squareroot expressions for $z_I^\pm$ in terms of the momenta. However, it can be simplified significantly  
when the L and R subamplitudes are both local (i.e.~do not themselves have any poles, meaning that they are matrix elements of local interactions); this is the case for all applications in this paper. In that case, Cauchy's theorem can be used to rewrite \reef{softsubrecrel1} as
\be
  \label{softrecsimple}
  A_n = \sum_{I, h_I} \bigg(
      \frac{\hat{A}^\text{L}_n(0)\hat{A}^\text{R}_n(0)}{P_I^2}
      + \sum_{i=1}^n \text{Res}_{z=1/a_i}
      \frac{\hat{A}^\text{L}_n(z)\hat{A}^\text{R}_n(z)}{z F(z) \hat{P}_I^2}
  \bigg)\,.
\ee
Again, this form assumes all L and R subamplitudes to be polynomial in the kinematic variables. We use it in this paper for L and R input which are either 4- or 5-point local matrix elements. 

The validity of the soft subtracted recursion relations rely on the absence of a pole at $z=\infty$. A  sufficient criterion for this was derived in \cite{Elvang:2018dco}. For the case of factorization into two subamplitudes with couplings of mass-dimension $[g_L]$ and $[g_R]$, respectively, the validity criterion of \cite{Elvang:2018dco} is simply
\be
  \label{validity}
  4-n - [g_L] - [g_R] - \sum_{i=1}^n s_i - \sum_{i=1}^n \sigma_i < 0 \,, 
\ee
where $s_i$ is the spin (not helicity) of the $i^\text{th}$ particle and $\sigma_i$ its soft weight.

%%%%%%%%%%%%%%%%%%%%%%%%%
\subsection{Soft Bootstrap}
%%%%%%%%%%%%%%%%%%%%%%%%%

The soft bootstrap uses the soft subtracted recursion relations as a method to explore the space of field theories, in particular EFTs of massless states with Adler zeros. The input are {\em assumptions} about the spectrum of the model and the soft weights of each species of particles as well as any global unbroken symmetry (such as supersymmetry). The lowest-point amplitudes (say at 4- and 5-point as relevant for our analysis) are then parameterized as local polynomials in the kinematic variables (we use the 4d spinor-brackets) in the most general form subject to the assumed properties and these then become the input for soft recursion. 

If the validity criterion \reef{validity} is {\em not} satisfied for a given higher-point amplitude, it means that there exists local interactions that trivially satisfy the softness constraints \cite{Elvang:2018dco}. Thus we are not able to restrict such cases. 

If the validity criterion \reef{validity} {\em is} satisfied, we use the recursion expression \reef{softrecsimple}: if there exists a theory with the desired proporties, then the LHS of \reef{softrecsimple} must yield a result independent of the shift parameters $a_i$. However, if the LHS of \reef{softrecsimple} has dependence on the $a_i$, the result is not a valid amplitude. What must have failed is the assumption about soft weights $\sigma_i$ of one or more states in the proposed model. This then rules out the existence of such a model.

Thus the soft recursion relations become a very effective tool for ruling out the existence of models: this is what is  called the soft bootstrap.  However, when a prospective model passes all available $a_i$-independence test, it is not proof of existence and further work is necessary to establish the model. 

We now provide some details of how the soft bootstrap was used in the maintext. 

%%%%%%%%%%%%%%%%%%%%%%%%%
\subsection{Example 1: Complex Scalar Quartic Galileon}
\label{app:n4nosusy}
%%%%%%%%%%%%%%%%%%%%%%%%%

Consider the complex scalar quartic Galileon in Section \ref{s:review}. The most general\footnote{Most general in the context of supersymmetry, which requires $A_4(Z Z Z Z)$, $A_4(Z Z Z \bar{Z})$, and their conjugates to be zero.} local 4-point amplitude at order $O(p^6)$ --- and hence coupling
mass-dimension $[g_4] = -6$ --- was given in \reef{preA4ZZbZZb} as
\be
\label{preA4ZZbZZbapp}
A_4(Z \bar{Z} Z \bar{Z}) = c_1 s t u + c_2 t^3\,.
\ee
Consider $A_6(Z \bar{Z}Z \bar{Z}Z \bar{Z})$.  
The soft recursion validity criterion \reef{validity} for the 6-scalar amplitude with two quartic Galileon vertices is 
\be
  0 > 4-6 +6+6 - 6\sigma_Z 
  =
  10 - 6\sigma_Z 
  ~~~\implies~~~
  \sigma_Z > 5/3\,.
\ee
Thus this amplitude is recursive only for $\sigma_Z=2$ (or greater). Let us apply the soft bootstrap with the assumption that $\sigma_Z=2$. There are 9 distinct  factorization channels of $A_6(Z \bar{Z}Z \bar{Z}Z \bar{Z})$. Using this in \reef{softrecsimple} one finds\footnote{See \cite{Elvang:2018dco} for details on how to solve the kinematic constraint \reef{phat} required by momentum conservation in both 4d and 3d kinematics.}  that the only case that passes the soft bootstrap (i.e.~yields a result independent of the $a_i$) is when $c_2 = 0$. This is the result stated in \reef{4ptsigmaZ2}. Assuming a further enhanced softness, $\sigma_Z \ge 3$, does not pass the soft bootstrap, so no such models exist.

Consider now $\sigma_Z=1$. If the real and imaginary parts of $Z$ to have softness 
\be   
  \label{sigma21}
  \sigma_\phi = 2\,,
  ~~~~
  \sigma_\chi = 1\,.
\ee
(option b of Section \ref{s:N1Gal}), then the validity criterion for $n=n_\phi+n_\chi=6$ gives 
\be
  0 > 4-6 +6+6 - 2 n_\phi - n_\chi 
  ~~~\implies~~~
  n_\chi < 2 \,.
\ee
The real-scalar 4-point amplitudes that follow from \reef{preA4ZZbZZbapp} are
\be
  A_4(\phi\phi\phi\phi) = A_4(\chi\chi\chi\chi)
  = \frac{3}{2} (c_1+c_2) s t u\,,
  ~~~~A_4(\phi\phi\chi\chi) =
  \frac{1}{2} \Big( c_1 s t u + c_2 (-s^3 + t^3+u^3) \Big)\,,
\ee
while $A_4(\phi\phi\phi\chi) = A_4(\phi\chi\chi\chi) = 0$.

Thus, there are not any non-trivial bootstrap tests because the 4-point amplitudes with a single $\phi$ vanish, hence there are not factorization channels for the 6-point amplitude with a single $\phi$.

%%%%%%%%

Instead we can test if this quartic 2-scalar Galileon with softness \reef{sigma21} is compatible with DBI. The complex scalar DBI amplitude is $A_4^\text{DBI}(Z \bar{Z} Z \bar{Z}) =  g t^2$ and from that it follows that
\be
  A_4^\text{DBI}(\phi\phi\phi\phi) = A_4^\text{DBI}(\chi\chi\chi\chi)
  = \frac{1}{2} g \big(s^2+t^2+u^2\big)\,,
  ~~~~A_4^\text{DBI}(\phi\phi\chi\chi) =
  \frac{1}{2}g  \big(-s^2 + t^2+u^2\big)\,,
\ee
and $A_4^\text{DBI}(\phi\phi\phi\chi) = A_4^\text{DBI}(\phi\chi\chi\chi) = 0$. These DBI amplitudes have soft weights 2 for both $\phi$ and $\chi$. 

For 6-point amplitudes with one vertex from DBI and one from the quartic Galileon, the validity criterion   gives $n_\chi < 4$, assuming softness \reef{sigma21}. This means we have a non-trivial test for $n_\chi=2$. This 6-point amplitude has two distinct factorization channels, one with a $\phi$ exchanged and one with a $\chi$ exchanged. For each case, we have to consider the DBI being on either side of the exhanged channel. All in all, this gives 20 diagrams in the recursive sum. One finds that the  bootstrap test is passed without constraints on $c_1$ and $c_2$, in accordance with the statement \reef{4ptsigmaZ1}.

%%%%%%%%%%%%%%%%%%%%%%%%%
\subsection{Example 2: $\mathcal{N}=2$ Quartic Galileon}
\label{app:n4susy}
%%%%%%%%%%%%%%%%%%%%%%%%%
In this example, we consider $\mathcal{N}=2$ Quartic Galileon from Section \ref{s:N2Gal4sigma1}. For a 6-point amplitude with $n_\phi$ and $n_\chi$ external scalars, $n_\lambda$ fermions, and $n_\gamma$ vectors and assumed  softness \reef{sigmas}, the validity condition \reef{validity} for factorization into two $\mathcal{N}=2$ 4-point Galileon subamplitudes is
\be
    n_\chi+\tfrac{1}{2} n_f+n_\gamma < 2\,,
\ee
using $n_\phi + n_\chi+ n_f+ n_\gamma = 6$. 
This means that apart from the all-$\phi$ Galileon amplitude, only the 2-fermion amplitude $A_4( \lambda^{1+} \bar\lambda ^{2-} \phi\phi)$ gives a non-trivial recursion. (The amplitudes $A_6(\chi \phi \phi\phi\phi)$ and $A_6(\gamma \phi \phi\phi\phi)$ trivially vanish because there are no available factorizations.) In the
 $\mathcal{N}=2$ model the amplitude
\be
\label{N2twofer}
A_4( \lambda^{1+} \bar\lambda ^{2-} \phi\phi) = \frac{1}{2\Lambda^6} (u^2-t^2) [13]\<23\> 
\ee
 is the same as in the $\mathcal{N}=1$ quartic Galileon, so this means at we are not testing the model beyond $\mathcal{N}=1$. 

The 3-brane effective action and the discussion of UV-completability from Section \ref{s:swampland} motivates  testing the Galileon model together with  DBI. Therefore, we consider  the three $\mathcal{N}=2$  models in the bullet points of Section \ref{s:N2Gal4sigma1}.

The validity criterion \reef{validity} at 6-point with input of one quartic vertex from (supersymmetric) DBI and one from the (supersymmetric) quartic Galileon gives
\be
  \label{valid1}
  0> 4-6+4+6- n_\gamma - \tfrac{1}{2} n_f - 2n_\phi - n_\chi - n_f 
   ~~~\implies~~~
    n_\chi+\tfrac{1}{2} n_f+n_\gamma < 4\,.
\ee
This implies that the soft subtracted recursion relations are valid for the 6-fermion amplitude $A_6(\lambda\bar{\lambda}\lambda\bar{\lambda}\lambda\bar{\lambda})$, which is the application described in Section \ref{s:N2Gal4sigma1}. There are 9 factorization channels and for each one of them there are two options for which side the DBI vertex sits. When adding up the contributions for the resulting 18 diagrams, carefully taking into account the sign for odd permutations of fermions, one finds the result stated in Section \ref{s:N2Gal4sigma1} that while the $\mathcal{N}=1$ quartic DBI-Galileon with $\sigma_Z=2$ passes the soft bootstrap, the $\mathcal{N}=2$ quartic DBI-Galileon with $\sigma_\phi=2$ and $\sigma_\chi=1$ does not. Note that at 6-point, the full power of the soft bootstrap is only accessed when done in 3d kinematics and that is what is used to obtain this exclusion result. We performed a similar test for $A_6(\phi\phi\lambda\bar{\lambda}\lambda\bar{\lambda})$ with the same outcome.

%%%%%%%%%%%%%%%%%%%%%%%%%
\subsection{Example 3: $\mathcal{N}=1,2$ Quintic Galileon}
\label{app:n5}
%%%%%%%%%%%%%%%%%%%%%%%%%

The results of soft bootstrap for the  $\mathcal{N}=1,2$ quintic Galileons are discussed in Sections \ref{s:N1Gal} and \ref{s:N2quinticsigmaZ1}. Here we provide additional details. 

\noindent {\bf $\mathcal{N}=1$ supersymmetry.} 
The analysis that leads to the 3-parameter family of  $\mathcal{N}=1$ compatible 5-point matrix elements is reviewed in Appendix \ref{app:N1quintic}. Via the supersymmetry Ward identity (and complex conjugation to the conjugate sector) these can all be written in terms of the 5-scalar amplitude \reef{A5scalar3par}. 

Consider first the 8-point bootstrap test with two 5-point vertices \reef{A5scalar3par} and softness \reef{sigmas}. The validity criterion gives that other than the all-$\phi$ amplitude the only amplitude that can be tested is the 2-fermion amplitude $A_8(\lambda\bar{\lambda}\phi\phi\phi\phi\phi\phi)$. It passes. 

Similarly, at 7-points with one 5-point vertex \reef{A5scalar3par} and one $\mathcal{N}=1$ quartic Galileon, only the all-$\phi$ amplitude and $A_7(\lambda\bar{\lambda}\phi\phi\phi\phi\phi)$ pass the validity criterion. It also passes. 

Moving on, we can get a lot more non-trivial tests at 7-point when one vertex is supersymmetric DBI. 
The validity criterion is 
\be
  0> 4-7+4+9- n_\gamma - \tfrac{1}{2} n_f - 2n_\phi - n_\chi - n_f 
      ~~~\implies~~~
    n_\chi+\tfrac{1}{2} n_f+n_\gamma < 4 \,.
\ee
This happens to be the same validity criterion as \reef{valid1}, but with $n=n_\gamma+ n_f+ n_\chi+n_\phi = 7$. 

First specialize to the $\mathcal{N}=1$ case where $n_\gamma=0$. We then have $n_\chi+\tfrac{1}{2} n_f < 4$, so that means the following amplitudes are soft constructible:
\be
  \label{eq:N1_tests}
  \begin{array}{lll}
  A_7(\phi\phi \phi \phi \phi\phi\phi)\,,
  &
  A_7(\phi\phi \phi \phi \phi\chi\chi)\,,
  &
  A_7(\phi\phi \phi \phi \chi\chi\chi)\,,
  \\
  A_7(\lambda^+\lambda^-\phi\phi\phi\phi \phi)\,,
  &
  A_7(\lambda^+\lambda^-\phi\phi\phi\phi \chi)\,,
  &
  A_7(\lambda^+\lambda^-\phi\phi\phi\chi \chi)\,,
  \\
  A_7(\lambda^+\lambda^-\lambda^+\lambda^-\phi\phi \phi)\,,
  &
  A_7(\lambda^+\lambda^-\lambda^+\lambda^-\phi\phi \chi)\,,
  &
  A_7(\lambda^+\lambda^-\lambda^+\lambda^-\lambda^+\lambda^- \phi)\,.
  \end{array}
\ee
(The amplitude $A_7(\phi\phi \phi \phi \phi\phi\chi)$ vanishes.)
The soft bootstrap is applied to each of these amplitudes and no constraints were found on the three parameters $q_1$, $q_2$, and $q_3$ of \reef{A5scalar3par}.

\noindent {\bf $\mathcal{N}=2$ supersymmetry.} 
A 2-parameter family \reef{N2susycond} of the $\mathcal{N}=1$ supersymmetric 5-point solution \reef{A5scalar3par} satisfy the 5-point $\mathcal{N}=2$ SUSY Ward identities. Setting $q_3 = 3 q_2$, 
we tested all constructible, non-trivial, and distinct 7-point amplitudes:
\be
  \label{eq:N2_tests}
  \begin{array}{ll}
  A_7(\phi \phi \phi \phi \phi \phi \phi)\,,
  &
  A_7(\phi \phi \phi \phi \phi \chi \chi)\,,
  \\
  A_7(\phi \phi \phi \phi \chi \chi \chi)\,,
  &
  A_7(\lambda^{1+} \lambda^{2-} \phi \phi \phi \phi \phi)\,,
  \\
  A_7(\lambda^{1+} \lambda^{2-} \phi \phi \phi \phi \chi)\,,
  &
  A_7(\lambda^{1+} \lambda^{2- }\phi \phi \phi \chi \chi)\,,
  \\
  A_7(\lambda^{1+} \lambda^{2-} \lambda^{1+} \lambda^{2-} \phi \phi \phi)\,,
  &
  A_7(\lambda^{1+} \lambda^{2-} \lambda^{2+} \lambda^{1-} \phi \phi \phi)\,,
  \\
  A_7(\lambda^{1+} \lambda^{2-} \lambda^{1+} \lambda^{2-} \phi \phi \chi)\,,
  &
  A_7(\lambda^{1+} \lambda^{2-} \lambda^{2+} \lambda^{1-} \phi \phi \chi)\,,
  \\
  A_7(\lambda^{1+} \lambda^{2-} \lambda^{1+} \lambda^{2-} \lambda^{1+} \lambda^{2-} \phi)\,,
  &
  A_7(\lambda^{1+} \lambda^{2-} \lambda^{1+} \lambda^{2-} \lambda^{2+} \lambda^{1-} \phi)\,,
  \\
  A_7(\gamma^+ \gamma^- \phi \phi \phi \phi \phi)\,,
  &
  A_7(\gamma^+ \gamma^- \phi \phi \phi \phi \chi)\,,
  \\
  A_7(\gamma^+\lambda^{1-}\lambda^{2-} \phi \phi \phi \phi)\,, 
  &
  A_7(\gamma^+ \lambda^{1-} \lambda^{2-} \phi \phi \phi \chi)\,, 
  \\
  A_7(\gamma^+ \lambda^{1-} \lambda^{2+} \lambda^{1-} \lambda^{2-} \phi \phi)\,,
  &
  A_7(\gamma^+ \lambda^{2-} \lambda^{1+} \lambda^{1-} \lambda^{2-} \phi \phi)\,,
  \\
  A_7(\gamma^+ \gamma^- \lambda^{1+} \lambda^{2-} \phi \phi \phi)\,.
\end{array}
\ee
(We are not listing amplitudes that are related to the above by conjugation.)
All pass the bootstrap without constraints on $q_1$ and $q_2$.

%%%%%%%%%%%%%%%%%%%%%%%%%%%%%%%%%%%%%%%
\section{$\mathcal{N}=1$ Quintic Galileon: $A_5(\bar{Z}Z\bar{Z} Z \bar{Z})$}
\label{app:N1quintic}
%%%%%%%%%%%%%%%%%%%%%%%%%%%%%%%%%%%%%%%
Here we briefly outline the analysis that leads to the 3-parameter solution \reef{A5scalar3par} for the $\mathcal{N}=1$ Quintic Galileon. The first step is to construct an ansatz for the local matrix elements $A_5(\bar{Z}Z\bar{Z} Z \bar{Z})$ and the independent ones related to it by $\mathcal{N}=1$ SUSY. In absence of 3-point interactions, these must be degree-8 polynomials in the spinor-helicity variables and they must respect little group scaling and Bose/Fermi symmetry. The table summarizes the number of such independent (under momentum conservation and Schouten identities) spinor-bracket polynomials for each of the given amplitudes:
\be
\label{3N1amps}
\begin{array}{lc}
\text{Amplitude} & \text{Number of indep.~terms}\\
 A_5(\bar{Z}Z\bar{Z} Z \bar{Z}) & 10 \\
 A_5(\bar{\lambda} \lambda \bar{Z} Z \bar{Z}) & 35 \\
 A_5(\bar{\lambda} \lambda \bar{\lambda} \lambda  \bar{Z}) & 16 \\
\end{array}
\ee
Each amplitude is now written as a linear combination of the respective spinor-helicity polynomials with arbitrary coefficients. This 61 parameter ansatz is subjected to the $\mathcal{N}=1$  SUSY Ward identities that relate the three amplitudes \reef{3N1amps} to each other (equivalent to eq.~(27) of \cite{Elvang:2017mdq}). A 6-parameter family results from this. 

Up to this point, no assumptions were made about the soft behavior of any of the particles. Imposing $\sigma_Z =\sigma_{\bar{Z}}  =1$ leaves 4 parameters unfixed. In order for the real part of $Z$ to be a Galileon $\phi$, we impose  $\sigma_\phi = 2$. This requires the conjugate amplitudes such as $A_5(Z\bar{Z}Z\bar{Z} Z)$ and we allow its 4 unfixed parameters to be unrelated to those of $A_5(\bar{Z}Z\bar{Z} Z \bar{Z})$. Now $\sigma_\phi = 2$ as well as the requirement that $A_5(\phi\phi\phi\phi\phi)$ is non-zero fixes the 4+4 parameters down to 3 parameters. 

The 3-parameter result for $A_5(\bar{Z}Z\bar{Z} Z \bar{Z})$ is the solution given in \reef{A5scalar3par}. In the above, it was not assumed that $\sigma_\lambda = 1$: this soft behavior emerged from the assumptions of $\sigma_\phi = 2$ and $\mathcal{N}=1$ supersymmetry. As described in Appendix \ref{app:n5} no further constraints arise from the soft boostrap performed at 7-points.

%%%%%%%%%%%%%%%%%%%%%%%%%%%%%
\section{Derivation of $\mathcal{N}=4$ SUSY Ward identities}
\label{app:SUSYWI}
%%%%%%%%%%%%%%%%%%%%%%%%%%%%%
We use the conventions for $\mathcal{N}=4$ supercharges and their action on the on-shell states from equation (4.31) of \cite{Elvang:2015rqa} (alternatively see eq.~(4.21) of \cite{Elvang:2013cua}). In short, $\tilde{Q}_A$ removes an upper $R$-index on a state whereas $Q^A$ adds $A$. Signs arise from odd-permutations of $R$-indices. 
In derivations of SUSY Ward identities, signs also occurs from interchanges of supercharges and fermionic states.

$\mathcal{N}=4$ supersymmetric DBI has full $SU(4)\times U(1)$ $R$-symmetry, but the quintic Galileon with its  5-scalar 
amplitude 
$A_5(Z^{12} Z^{34}Z^{12}Z^{34}Z^{12})$ can at most preserve 
$SU(2) \times SU(2)$. The 5-scalar amplitude and those related to it by $\mathcal{N}=4$ supersymmetry is in a sector that lies between MHV and NMHV: this means that it is MHV with respect to $Q^1$ and $Q^2$ as well as with   $\tilde{Q}_3$ and $\tilde{Q}_4$, while being NMHV with respect to the two other pairs. 

We now derive the Ward identity \reef{5ZN4} used in Section \ref{s:N4gGal5}. Start with
\be
  \begin{split}
  0 =&\, \big\< [\tilde{Q}_3, + - \lambda^3 - Z^{12}] \big\> \\
  =&\,
  |2\> \big\< + \lambda^{124} \lambda^3 - Z^{12} \big\>
  +|3\> \big\< + - + - Z^{12} \big\>
  -|4\> \big\< +- \lambda^3  \lambda^{124} Z^{12} \big\>.
  \end{split}
\ee
Project in $\<2|$ to find
\be
 \label{aSUSY1}
 \big\< +- \lambda^3  \lambda^{124} Z^{12} \big\>
  =
 \frac{\<23\>}{\<24\> } \big\< +- + - Z^{12} \big\>\,.
\ee
Next, consider 
\be
  \begin{split}
  0 =&\, \big\< [\tilde{Q}_4, + - Z^{34} \lambda^{124} Z^{12}] \big\> \\
  =&\,
  -|2\> \big\< + \lambda^{123} Z^{34} \lambda^{124} Z^{12} \big\>
  -|3\> \big\< + - \lambda^{3} \lambda^{124} Z^{12} \big\>
  +|4\> \big\< +- Z^{34} Z^{12}  Z^{12} \big\>.
  \end{split}
\ee
Project in $\<2|$ to find
\be
 \label{bSUSY1}
 \big\< +- Z^{34} Z^{12}  Z^{12} \big\>
  =
  \frac{\<23\>}{\<24\> } 
  \big\< +- \lambda^3  \lambda^{124} Z^{12} \big\>\,.
\ee
Combine \reef{aSUSY1} and \reef{bSUSY1} to get
\be
 \label{cSUSY1}
 \big\< +- Z^{34} Z^{12}  Z^{12} \big\>
  =
  \frac{\<23\>^2}{\<24\>^2} 
  \big\< +- + - Z^{12} \big\>\,.
\ee
Now use $\tilde{Q}_3$ and $\tilde{Q}_4$ again but this time to convert the remaining pair of vectors to a pair of scalars. The result is 
\be
 \label{dSUSY1}
 \big\< Z^{34} Z^{12} Z^{34} Z^{12}  Z^{12} \big\>
  =
  \frac{\<13\>^2}{\<23\>^2} 
  \big\< +- Z^{34} Z^{12}  Z^{12} \big\>\,.
\ee
Together with \reef{cSUSY1}, this then gives the SUSY Ward identity
\be
 \label{eSUSY1}
 \big\< Z^{34} Z^{12} Z^{34} Z^{12}  Z^{12} \big\>
  =
  \frac{\<13\>^2}{\<24\>^2} 
  \big\< +- + -  Z^{12} \big\>\,.
\ee

Doing the equivalent calculation with $Q^1$ and $Q^2$ leads to the Ward identity
\be
 \label{fSUSY1}
 \big\< Z^{12}Z^{34} Z^{12} Z^{34} Z^{12}   \big\>
  =
  -\frac{[24]^2}{[13]^2} 
  \big\< +- + -  Z^{12} \big\>\,.
\ee
It the follows from \reef{eSUSY1} and \reef{fSUSY1} that 
\be
  \label{gsusy1}
  \big\< Z^{34} Z^{12} Z^{34} Z^{12}  Z^{12} \big\>
  =
  -\frac{s_{13}}{s_{24}} 
  \big\< Z^{12}Z^{34} Z^{12} Z^{34} Z^{12}  \big\>\,.
\ee
Identifying $Z=Z^{12}$ and $\bar{Z}=Z^{34}$, the $\mathcal{N}=4$ SUSY Ward identity \reef{5ZN4} used in the main text follows from \reef{gsusy1}.

Note that the relations \reef{eSUSY1} and \reef{fSUSY1} are individually examples of $\mathcal{N}=2$ SUSY Ward identities. For example, \reef{fSUSY1} is the conjugate of the $\mathcal{N}=2$ identity  \reef{A5SUSYWI1}
that was derived from the  $\mathcal{N}=2$  superamplitude \reef{superA5}. However, when combined, all four supercharges were used so \reef{gsusy1} is truly an $\mathcal{N}=4$ SUSY Ward identity.

%%%%%%%%%%%%%%%%%%%%%%%%%%%%%%%%%%%%%%%
%\printbibliography

\bibliographystyle{utphys}
\bibliography{extSUSYGal.bib}
%\bibliographystyle{JHEP}

%%%%%%%%%%%%%%%%%%%%%%%%%%%%%%%%%%%%%%%
\end{document}